\documentclass[aps,prb,twocolumn,showpacs]{revtex4}
\usepackage{graphicx}%
\usepackage{dcolumn}
\usepackage{amsmath}
\usepackage{latexsym}
\usepackage[colorlinks,citecolor={blue},linkcolor={blue}]{hyperref}
\newcommand{\dg}{\dagger}
\newcommand{\up}{\uparrow}
\newcommand{\dwn}{\downarrow}
\newcommand{\pmbf}{\boldsymbol}
\newcommand{\dl}{\delta_0}
\begin{document}
\title{Adiabatic charge and spin transport in interacting quantum wires}
\author{Prashant Sharma}
\author{ Claudio Chamon}
\affiliation{Department of Physics, Boston University, Boston, MA
  02215}
\date{\today}
\begin{abstract}
  We study charge and spin transport through an interacting quantum
  wire, caused by backscattering off an effective impurity potential
  with a periodic time-dependence.  The adiabatic regime of this pump
  for charge and spin is shown to depend on the presence of
  interactions in the wire. For the non-interacting case the
  transported quantities in a period are calculated and found to be
  adiabatic (independent of pumping frequency) for all frequencies
  $\Omega\ll v_F/\ell$, where $\ell$ is the range of the scattering
  potential, and $v_F$ is the Fermi velocity; this result is along the
  lines of adiabatic pumping in mesoscopic systems. On the other hand,
  we show that for a wire with repulsive electron-electron
  interactions the adiabatic regime is confined to $\hbar\Omega\ll
  \omega_\Gamma$, with $\omega_\Gamma$ an energy scale set by the
  backscattering potential. Using symmetry and scaling properties of
  the quantum wire Hamiltonian we relate the charge $Q$ and spin $S$
  transported through the wire in a period $2\pi/\Omega$ to an
  integral involving quasi-static backscattering conductances.  We
  also show that the pumped charge $Q$ (or the spin $S$) is quantized
  in the adiabatic limit if the conductance of the system is zero at
  the stable fixed point of the renormalization-group (RG)
  transformations. The quanta transported are given by the winding
  number of a complex coupling $\Gamma(\Omega t)$ as it traverses a
  closed path ${\cal C}$ enclosing the origin $\Gamma=0$. No adiabatic
  charge or spin is transported when the RG fixed point corresponds to
  a perfectly conducting wire, or if the path ${\cal C}$ does {\it
    not} enclose the point $\Gamma=0$. By contrast, for a RG marginal
  conductance -- which is the case for non-interacting electrons --
  the charge transported in a cycle is not quantized, rather it is
  proportional to the area enclosed by the path or circuit ${\cal C}$,
  in agreement with previous studies of non-interacting mesoscopic
  systems. Finite size, temperature, and frequency effects on the
  transported quantities follow from the relation between adiabatic
  transport coefficients (backscattering conductances) and the
  quasi-static conductance of the system.
\end{abstract}

 \maketitle
\section{Introduction}
\label{intro}
Quantum adiabatic charge transport through a system is achieved by
a slow periodic modulation in time of a set of external
potentials. This phenomenon has been investigated in
one-dimensional non-interacting systems with gapped excitations
(band insulators) \cite{thouless83,niu86,niu90}, for which case
the pumped charge $Q$ (charge transported through the system in a
single period) was shown to be quantized. In the context of
adiabatic transport through quantum dots, recent theoretical work
on pumping \cite{brouwer98,aleiner98,zhou99} finds no such
universal behavior for $Q$, which strongly depends on how the pump
is operated. On the other hand, for gapless systems in one space
dimension (1D Luttinger liquids), it is possible to pump weakly
quantized charge and spin by the temporal variation of a spatially
localized potential~\cite{sharma01}.

In our earlier work, Ref.~\onlinecite{sharma01}, the weak quantization
of adiabatically pumped charge or spin in a Luttinger liquid was
argued to follow from the repulsive nature of electron-electron
interactions. These arguments were given by solving exactly the
special case of the spinless Luttinger model with interaction
parameter $K=1/2$ in the presence of an impurity with arbitrary
time-dependence $\Gamma(\Omega t)$. Quantization of pumped charge was
shown in the limit $\Omega\to 0$ with power law corrections $\Delta
Q\sim\left(\Omega/\omega_\Gamma\right)^2$ determined by the barrier
energy scale $\omega_\Gamma$. The quanta transported are given by the
winding number of the complex backscattering amplitude $\Gamma$.
Therefore, the pumped quantities are invariant under (small)
continuous deformations of the path ${\cal C}$, around the origin of
the complex plane, traced by $\Gamma$.

In this paper, we prove the adiabatic quantization of the pumped
charge and spin for gapless interacting systems using the properties
of chiral symmetry and scale invariance. We derive a connection
between the adiabatically pumped quantities and an integral involving
the quasistatic (equilibrium) value of the conductance through the
system. The quasistatic conductance is calculated using an
instantaneous renormalized (because of bulk interactions)
backscattering amplitude $\tilde\Gamma(t)$.  For example, in its
simplest form for a spinless system, the expression for the pumped
charge is:
\begin{equation}\label{qb-dccond-intro}
Q=\frac{1}{2\pi i}\oint \frac{d\Gamma}{\Gamma}
\left[1- 2\pi G_{\tilde\Gamma}\right],
\end{equation}
where $G_{\tilde\Gamma}$ is the equilibrium conductance of the
one-dimensional wire with the instantaneous (renormalized)
backscattering amplitude $\tilde\Gamma(t)$. Our derivation generically
links quantization of adiabatic charge transport to insulating
renormalization group (RG) fixed points of the interacting wire with
an impurity. Thus, whenever the backscattering potential leads to
insulating behavior ($G_{\tilde\Gamma}\to 0$) in response to an
applied DC voltage, the pumped charge is perfectly quantized, albeit
the time to transport such charge adiabatically is large (we discuss
in this paper the frequency scales for adiabaticity). The quanta is
given by the winding number (about the origin $\Gamma=0$) of the
backscattering amplitude $\Gamma$. For the particular case of
non-interacting electrons, for which the backscattering amplitude is
marginal in the RG sense, the aforementioned quantization is lost, and
we show that our result recovers the ``area law'' relation obtained by
Brouwer \cite{brouwer98}.

We emphasize that the relation between the adiabatic charge and spin
transported in a periodic cycle and the quasistatic conductance is
rather general; it applies to systems with both short and long-range
bulk interactions, and is derived here without taking recourse to
bosonization. Within the bosonization approach, arguments have been
put forth in Ref.~\onlinecite{feldman-comment} that the quantization
is a consequence of pinning the bosonic field at the impurity to a
minimum of a single cosine potential when the backscattering potential
is relevant; the displacement of the bottom of the potential after a
complete cycle leads to the quantization condition. Here we show that
the mechanism of pumping is based on controlling the spatially
localized backscattering process that breaks a continuous global
(chiral) symmetry of the bulk one-dimensional system, and point out
the crucial role of interactions in the wire in determining the
transported quantities.

The rest of this paper is organized as follows. In Section II we
consider the case of charge pumping through a quantum wire of
interacting spinless fermions with gapless excitations. We introduce
the model in Section IIA and relate it to a time dependent single
impurity model, with a (complex) coupling $\Gamma(t)$. In Section IIB
we show how the temporal variation of the phase of $\Gamma$ can lead
to the transport of charge $Q$ through an otherwise clean quantum
wire. This is done by making use of the Ward identities which are
derived in that section. In Section IIC we obtain a relation between
the pumped charge in a cycle and the backscattering-current
conductance of the wire with renormalized parameter.
In Section IID we relate the pumped quanta to the usual conductance of the quantum
wire.  In Section III we consider electrons with spin and derive
expressions for the spin ($S$) and charge ($Q$) transported per cycle by
the spin and charge pumps.  These expressions allow us to look at the
limits of quantization and at the effect of finite temperature and
finite size on the value of $Q$ and $S$, which is the content of
Section IV. In Section V we solve the time-dependent impurity problem
exactly for non-interacting electrons, and also present details of the
calculations used in our earlier publication,
Ref.~\onlinecite{sharma01}, for the spinless Luttinger model at the
$K=1/2$ point. We identify the adiabatic regime, and compare the
results with the approach of Sections II and III.
Finally, in Section VI we conclude with a summary of our main
contributions.

\section{Transport of spinless electrons by parametric pumping}

\subsection{Model Hamiltonian}
\label{cp-model}

An adequate low energy description of a system of 1D fermions is
obtained by separating the Fermi field $\psi(x)$ into right (R) and
left (L) moving chiral fields near the Fermi points $\pm k_F$
\begin{equation}
\psi(x,t)=e^{ik_F x}\psi_R(x,t) + e^{-ik_F x}\psi_L(x,t).
\label{chiral}
\end{equation}
This turns a non-relativistic action into a relativistic Dirac fermion
action in (1+1) space-time dimensions, with the Fermi velocity $v_F$
playing the role of the speed of light \cite{stone}.  Our notation
assumes spinless fermions although, as we show later, it can be
straightforwardly generalized to include spin.  We consider such a
system with a Hamiltonian ${H}$ that preserves the global chiral
symmetry \footnote{We would like to emphasize that a strictly linear
  spectrum is not a necessary requisite for determining the nature of
  the adiabatic limit. This is because the adiabatic limit for a
  1D system with global chiral symmetry is determined by a set of
  relevant operators, and terms that contribute to non-linearity of
  the spectrum are irrelevant in the RG sense.\cite{haldane81}}
\begin{eqnarray}
\Psi(x)\to e^{i\alpha\gamma}\Psi(x),
\nonumber \\
\bar\Psi(x)\to \bar\Psi(x)e^{i\alpha\gamma},
\label{chiral-trans}
\end{eqnarray}
where
\begin{eqnarray}
\Psi=\left(
\begin{array}{c}\psi_R \cr {\psi_L} \cr \end{array}
\right),\;\mbox{and}\;\bar\Psi=\Psi^{\dg}\gamma^0=(\psi^{\dg}_L,\psi^{\dg}_R).
\label{psi-rl}
\end{eqnarray}
Here the $\gamma$-matrices are $\gamma^{0,1}=\sigma_{x,y}$, and
$\gamma=i\gamma^{0}\gamma^{1}=-\sigma_{z}$, with $\sigma_{x,y,z}$
being the Pauli matrices; and we choose $v_F=e=\hbar=1$.
The chiral symmetry ensures that the axial charge $\int dx\;\tilde j=\int
dx\;\bar\Psi\gamma^{0}\gamma\Psi$ is conserved in the absence of an external
electromagnetic field (``chiral anomaly'').

Adding a short range impurity potential $V(x)$ -- non-zero only in a
finite range of length $\ell\ll L$, the length of the wire -- gives rise to
an additional term in the Hamiltonian ${H}$:
\begin{equation}
\begin{split}
H_{\rm imp}&=\int dx\; V(x)\psi^{\dg}(x)\psi(x)\\
&=\int dx\; V(x)e^{i2k_Fx}\psi_L^{\dg}(x)\psi_R(x) +H.c.\\
&+\int dx\; V(x)\left[\psi_R^{\dg}(x)\psi_R(x) + R\to L\right] .
\end{split}
\end{equation}
Here the first term describes the backscattering of right movers into
left movers and vice versa, while the second term describes forward
scattering of both right and left movers with identical scattering
amplitude and phase shifts. The idea of parametric pumping in a
quantum wire \cite{sharma01} is to generate a current by varying
parameters that control the ``shape'' of this scattering potential.
Since the current (axial charge $\tilde j$) involves the difference in
the number of right and left movers in the wire, the forward
scattering term, which does not distinguish between the right and left
movers, plays no role in generating this current.  It is only the
backscattering terms which can lead to a non-zero current when the
potential $V(x)$ is suitably manipulated. Thus, for the purpose of
determining the pumped current at low energies (when the
continuum field theory description (\ref{psi-rl}) holds), we can write
down an effective Hamiltonian which describes the most relevant (in
the RG sense) backscattering processes. Thus the impurity contribution
to $H$ can be written as a local ($x=0$) backscattering term:
\begin{eqnarray}
H_{\Gamma} &=&
\bar\Psi(0)
\left[\begin{array}{cr}
{\Gamma}  & 0 \cr 0 & {\Gamma^*}\cr\end{array}\right] \Psi(0)
\label{himp1} \\
 &\equiv&\Gamma {\cal Q} + {\cal Q}^{\dg}\Gamma^{*}.
\label{himp}
\end{eqnarray}
Here ${\cal Q}$ represents the local operator $\psi^{\dg}_L\psi_R(0)$
and $\Gamma=\int dx\;V(x)e^{i2k_Fx}$ represents the $2k_F$-
Fourier-transform of the impurity potential. An important point about
this mapping from a local potential $V(x)$ on to a delta function
potential $\Gamma$, is that it is valid only at energies lower than
$\Lambda_\ell\equiv v_F/\ell$ where $\ell$ is a measure of the range of the
potential\footnote{Note that this does {\it not} mean that the
  continuum field theory itself has a cut-off $v_F/\ell$, but only that
  at higher energies the potential structure gets resolved.}. Let us
note here that by simple dimension counting (in natural units, with
$\hbar=e=v_F=1$) the coupling $\Gamma$ has {\it zero dimension}.
\begin{figure}
\includegraphics[scale=.35]{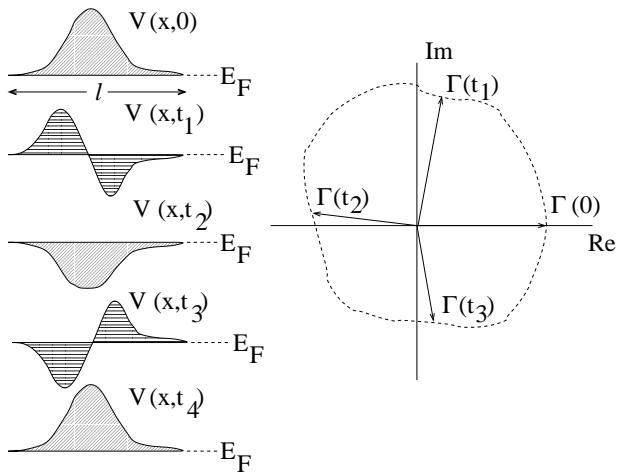} \caption{Variation of
$\Gamma$ due to $V(x,t)$ changing shape by changing its value
about $E_F$.} \label{gammat}
\end{figure}

If we vary the ``shape'' of the potential $V(x)$ in time, then it
amounts to varying both the phase $\phi$ and amplitude of $\Gamma$ in
time.  This can be understood qualitatively as follows (see also
Fig~\ref{gammat}). Consider a time ($t=0$ in Fig~\ref{gammat}) when
the potential profile is symmetric about $x=0$, then the Fourier
transform is real and the phase $\phi=0$. At a later time, when the
profile is anti-symmetric, the phase is $\phi=\pi/2$, and then, as the
potential changes sign, $\phi=\pi$ until the potential regains its
original shape and the phase goes to $\phi=2\pi$.  Thus, by changing
the potential profile, one gets an effective time dependent delta
function potential $|\Gamma(t)|e^{i\phi(t)}\;\delta(x)$.
\subsubsection{Double barriers: a realization of $\Gamma(t)$}

In the pumping device proposed in Ref.~\onlinecite{sharma01}, the
changing $V(x,t)$ of Fig.~\ref{gammat} is realized by oscillating two
potential barriers placed some distance apart:
\begin{equation}
\begin{split}
V(x,t)=V_0^-(x)f_-(t)+V_0^+(x)f_+(t),
\end{split}
\end{equation}
with out of phase periodic functions $f_+$ and $f_-$ which change sign
in a cycle.  For example, $f_-(t)=\cos\omega t$ and
$f_+(t)=\cos(\omega t+\theta)$.  We identify
\begin{equation}
\begin{split}
\Gamma(t)=\tilde V^-_0(2k_F)f_-(t)+\tilde V^+_0(2k_F)f_+(t),\,{\rm where}\\
\tilde V_0^{-,+}(2k_F)=\int dx\; e^{i2k_Fx}\;V^{-,+}_0(x).
\end{split}
\end{equation}
If $V_0^{-,+}(x)$ is asymmetrical about $x=0$, then the Fourier
transforms $\tilde V_0^{-,+}$ are not real and as a result
$\Gamma(t)$ is complex. Furthermore, if the functions are
different so that ${\rm Arg}[\tilde V_0^-]\neq{\rm Arg}[\tilde
V_0^+]$ then by varying the amplitudes and the signs of the
complex quantities $f_{-(+)}\tilde V_0^{-(+)}$ (accomplished by
varying $f_{\pm}$) one can reach any point in the complex $\Gamma$
plane. Thus, by choosing appropriate functional forms
$f_{\pm}(t)$, one can trace any desired path in the complex
$\Gamma$ plane (as in Fig.~\ref{gammat}).

\subsubsection{Special case: simple form for $V(x,t)$}
\label{special-form}

We choose delta function forms for $V_0^-(x)$ and $V_0^+(x)$:
$$
V_0^-(x)\equiv\gamma^-\delta(x+\ell/2),\,\,
V_0^+(x)\equiv\gamma^+\delta(x-\ell/2)
$$
 to obtain:
$$
V(x,t)=\gamma^-\delta(x+\ell/2)\cos\omega t +
\gamma^+\delta(x-\ell/2)\cos(\omega t +\theta).
$$
This gives an effective coupling
$$
\Gamma=\gamma^-e^{-ik_F\ell}\cos\omega t + \gamma^+
e^{ik_F\ell}\cos(\omega t +\theta),
$$
which is of the form $\Gamma(t)\equiv \Gamma_{+\omega}e^{i\omega
t}+\Gamma_{-\omega}e^{-i\omega t}$, where
$\Gamma_{\pm\omega}=V_0^+e^{-ik_F \ell} +V_0^-e^{ik_F\ell\pm i\theta}$.
Written in this way, the time dependence of $\Gamma$ can be
distinguished from that obtained by applying a voltage -- when only
{\it one} of $\Gamma_{\pm\omega}$ is present, as in
Ref.~\onlinecite{chamon95}.

The phase $\phi$ of the complex valued parameter $\Gamma$ is time
dependent whenever $2k_F\ell\neq n\pi$ and $\theta\neq n\pi$ ($n=0,\pm
1\ldots$).  The first condition is the off-resonance condition,
essential for backscattering to occur, while the second is essential
for traversing a closed path in the $\Gamma$ plane with a non-zero
area. The rate of change of this phase, $\dot\phi$, can be written as:
$$
\dot\phi(t)=\frac{\omega \gamma^- \gamma^+\sin\theta\sin 2k_F\ell}{X^2(t)},
$$
where
\begin{eqnarray*}
 X^2(t)&=& \cos^2k_F\ell\left[\gamma^+\cos(\omega t+\theta)
+\gamma^-\cos\omega t\right]^2 \\ &+&
\sin^2k_F\ell\left[\gamma^+\cos(\omega t+\theta) -\gamma^-\cos\omega
t\right]^2.
\end{eqnarray*}
It is clear from this expression that $\dot\phi(t)$ does not change
sign with time, and thereby takes the complex quantity $\Gamma$ around
a contour, as shown in Fig~\ref{gammat}.

\subsection{Pumping parameter for the charge pump}
\label{chargepump}

In the rest of this section we will show that the phase $\phi$ of the
complex quantity $\Gamma$ is the pumping parameter. The total
Hamiltonian is:
\begin{equation}
{\cal H}=H+H_{\Gamma}.
\label{hamiltonian}
\end{equation}
The backscattering term
(\ref{himp}) breaks the continuous chiral symmetry
(\ref{chiral-trans}), and thus gives rise to the backscattering
current
\begin{eqnarray}
 J_{\Gamma}(t) &=& -\frac{1}{2}\int dx\;\partial_t  \tilde j(t,x)
\label{jgamma} \\
&=&\frac{i}{2}\int\! dx\; [\tilde j (t,x),H_{\Gamma} ]
= i\Gamma{\cal Q}(t) + {\rm H.c.},
\label{jgamma-q}
\end{eqnarray}
which measures the rate of change of total charge of the right movers
in the system due to the backward scattering of the chiral fermions.
It is important to note that we do not consider, in ${\cal H}$, the
forward scattering term arising from the impurity potential. This is
because forward scattering involves local fermion density and
therefore conserves the chiral charge.

In order to look at the equilibrium expectation value of
(\ref{jgamma}) we consider the functional
\begin{equation}
Z[\Gamma]=\langle S_{\Gamma}(\infty,-\infty) \rangle,
\end{equation}
with a constant (other then the adiabatic switching on and off)
source term $H_{\Gamma}$ that connects the vacuum state of $H$ at
times $t=-\infty$ with the same state at $t=\infty$. As a
consequence of the global chiral symmetry a change of variables,
as in (\ref{chiral-trans}), with an infinitesimal $\alpha$ leaves
the Hamiltonian ${H}$ invariant. The only change comes from the
source term: $H_{\Gamma}\to H_{\Gamma}+\alpha J_{\Gamma}$.  This
implies:
\begin{equation}
0=\delta Z[\Gamma]
=-i\alpha\int_{-\infty}^{\infty}dt\;\langle J_{\Gamma}(t)\rangle_{\Gamma}
\end{equation}
where $\langle\cdots\rangle_{\Gamma}$ denotes the expectation value in
equilibrium with a non-zero  $\Gamma$. Since an equilibrium
expectation value should be time-independent, it follows that
$\langle J_{\Gamma}(t)\rangle_{\Gamma}=0$.

When $\Gamma\to \Gamma(t)$ the Hamiltonian ${\cal H} \to {\cal
  H}[\Gamma(t)]$, and the system is driven out of equilibrium.  Then
the instantaneous backscattering current can be non-zero. In order to
calculate this current we introduce the generating functional for a
closed time path (CTP) \cite{schwinger61,rammer86}:
\begin{eqnarray}
{\cal Z}[\Gamma_{+},\Gamma_{-}]=\langle 0| S_{\Gamma_{-}}(-\infty,\infty)
S_{\Gamma_{+}}(\infty,-\infty) |0\rangle,
\end{eqnarray}
where $|0\rangle$ is the state at time $t=-\infty$ when the source $\Gamma(t)$ is
adiabatically switched on. The $S$-matrix
\begin{eqnarray}
S_{\Gamma_{+}}(\infty,-\infty)=T{\rm exp}\left[-i\int_{-\infty}^{\infty}dt\;\Big\{{H}+
H_{\Gamma_{+}}(t)\Big\}\right]\nonumber
\end{eqnarray}
evolves the state $|0\rangle$ forward in time with a source
$\Gamma_{+}(t)$, while $S_{\Gamma_{-}}(-\infty,\infty)$ evolves this
state backward in time with a source $\Gamma_{-}(t)$.  Note that
choosing $\Gamma_{+}=\Gamma_{-}$ for all times makes ${\cal
  Z}[\Gamma]=1$.  We also note here that our choice of a single
initial state is valid only for the zero temperature formalism. In
general a non-zero temperature (or any other initial mixture of
states) can be straightforwardly accommodated by including an
additional $S$-matrix operator $S(-\infty-i\beta,-\infty)$ and tracing
over all initial states \cite{rammer86}.
\subsubsection{Ward-Takahashi Identities}
Because of the global chiral symmetry of ${H}$, not all Green
functions generated by the functional ${\cal Z}[\Gamma_{\pm}]$ are
independent. General identities between Green functions of different
orders of the local operator ${\cal Q}$ can be derived by imposing the
constraint that the generating functional, and any expectation value
derived from it, is invariant under the global chiral transformation
(\ref{chiral-trans}). This implies that:
\begin{eqnarray}
0=\frac{\delta}{i\delta\alpha}\Big\langle{\cal Q}(t)\Big\rangle=
\Big\langle{\cal Q}(t)\Big\rangle - \int_{-\infty}^{t}\!dt'
\Big\langle\left[{\cal Q}(t) , J_{\Gamma}(t')\right]\Big\rangle,\,\,\,\,\,
\label{wt-id-1}
\end{eqnarray}
where the angular brackets denote expectation value of the Heisenberg
operator at time $t$, and $J_{\Gamma}(\tau)= i\Gamma(\tau){\cal
Q}(\tau)+{\rm H.c.}$. Note that there is no restriction on the time
dependence of $\Gamma(t)$, so that these relations constitute
non-equilibrium generalizations of the Ward-Takahashi (WT) identities
derived in equilibrium field theory \cite{zinnjustin}.  In terms of
operators defined earlier, the WT identities are:
\begin{subequations}
\label{wi}
\begin{eqnarray}
\Big\langle H_{\Gamma}(t)\Big\rangle =-i\int_{-\infty}^{t}dt'\Big\langle
[J_{\Gamma}(t) , J_{\Gamma}(t')]\Big\rangle \label{wi-1}\\
\Big\langle J_{\Gamma}(t)\Big\rangle = i\int_{-\infty}^{t}dt'\Big\langle
[H_{\Gamma}(t) , J_{\Gamma}(t')]\Big\rangle. \label{wi-2}
\end{eqnarray}
\end{subequations}
%

\subsubsection{Pumped charge as a consequence of phase change of $\Gamma$}

In order to calculate the pumped charge let us consider a time
interval $[t_0,t_1]$, with $t_0< t_1$, in which the pumping parameter
can be written as $\Gamma(t)=\Gamma_0(t)+\delta\Gamma(t)$, where
$\delta\Gamma(t)$ vanishes smoothly outside the interval. We can now
write our Hamiltonian as ${\cal H}_0+\delta{\cal H}_{\Gamma}$, where
$\delta{\cal H}_{\Gamma}$ vanishes with the perturbation
$\delta\Gamma(t)$ outside the chosen time interval.  Then the current
generated by backscattering can be written in the interaction picture
with respect to the Hamiltonian ${\cal H}_0$:
\begin{equation}\label{ip-t}
I(t)=\left\langle S^{\dg}(t,t_0)\,J_{\Gamma}(t)\,S(t,t_0)\right\rangle,
\end{equation}
where all operators are in the interaction picture with respect to
${\cal H}_0$, and $J_{\Gamma}(t)= i\Gamma(t){\cal Q}(t)+{\rm H.c.}$.
The $S$-matrix \cite{rammer86}
in the interaction picture is:
\begin{equation}\label{u-int}
\begin{split}
S(t,t_0) =T e^{-i\int_{t_0}^{t}dt'\;\delta H_{\Gamma}(t')}.
\end{split}
\end{equation}
Here $T$ stands for time ordering, and $\delta H_{\Gamma}(t)$ is the
time-dependent perturbation in the interaction representation.

As a consequence of global chiral invariance, when the pumping
parameter $\Gamma(t)$ has a constant phase $\phi_0$ in the time
interval $[t_i,t_f]$ (where the system is in an equilibrium state for
times $t<t_i$) then the pumped charge in that interval
\begin{equation}\label{const-phi}
i\frac{\partial}{\partial\phi^{(+)}_0}Z[\Gamma^{(+)},\Gamma^{(-)}]
=\int_{t_i}^{t_f}dt\Big\langle J_{\Gamma_0}(t)\Big\rangle =0,
\end{equation}
Therefore, a phase change is the single important pumping parameter.
Notice that in order to achieve a time-dependent phase we need to vary
a minimum of {\it two} experimental parameters.

\subsection{Charge pump: backscattering conductance}
\label{cpump-gb}

Consider now a change in the phase of $\Gamma(t)$ as a perturbation:
\begin{subequations}
\begin{equation}
\begin{split}
\Gamma(t)&=|\Gamma(t)|e^{i\phi(t)},\\
\delta\phi(t)&=\phi(t)-\phi(t_0), \;{\rm and}\\
\Gamma_0(t)&=|\Gamma(t)|e^{i\phi(t_0)}.
\end{split}
\end{equation}
Then, the perturbation parameter can be explicitly written as:
\begin{equation}
\delta\Gamma(t)=\Gamma_0(t)\Big[\cos\delta\phi(t)+i\sin\delta\phi(t)-1\Big].
\end{equation}
\end{subequations}
Therefore, for a small enough change $\delta\phi(t)$ in the interval
$[t_0,t_1]$, we can evaluate (\ref{ip-t}) perturbatively to first
order in $\delta\phi(t)$:
\begin{subequations}
\label{delta-ip}
\begin{equation}\label{delta-ip1}
\begin{split}
  I(t)& = \Big\langle J_{\Gamma_0}(t)\Big\rangle
  - \delta\phi(t)\Big\langle{H_{\Gamma_0}}(t)\Big\rangle\\
  &- i \int_{t_0}^{t}dt'\delta\phi(t')\Big\langle[J_{\Gamma_0}(t),
  J_{\Gamma_0}(t')]\Big\rangle ,
\end{split}
\end{equation}
where all operators are in the Heisenberg representation with respect
to the unperturbed Hamiltonian ${\cal H}[\Gamma_0]$.

Using the WT identities (\ref{wi-1})-(\ref{wi-2}) in (\ref{delta-ip})
we obtain:
\begin{equation}\label{ip0}
\begin{split}
  I(t) \simeq &\Big\langle J_{\Gamma_0}(t)\Big\rangle +
  i\int_{t_0}^{t} dt'\;\delta\dot\phi(t')\\
  &\times\int_{-\infty}^{t'}\!\! dt'' \Big\langle[J_{\Gamma_0}(t),
  J_{\Gamma_0}(t'')]\Big\rangle,
\end{split}
\end{equation}
\end{subequations}
The pumped charge in a small time interval $[t_0,t_1]$ can be written
as:
\begin{equation}\label{dq0}
\begin{split}
  \delta Q(t_1;t_0) = & i\int_{t_0}^{t_1}dt\int_{t_0}^{t}
  dt'\;\delta\dot\phi(t')\\ &\times\int_{-\infty}^{t'}\!\! dt''
  \Big\langle[J_{\Gamma_0}(t), J_{\Gamma_0}(t'')]\Big\rangle,
\end{split}
\end{equation}
where the first terms contribution vanishes due to (\ref{const-phi}).
From the above expression we see that the role of WT identities is to
ensure that a non-zero pumped charge is a consequence of temporal
variation in the phase of the backscattering amplitude. We note that
in obtaining Eq.~(\ref{dq0}) we have explicitly shown that $\phi(t)$
is the pumping parameter.

\begin{figure}
\includegraphics[scale=.35]{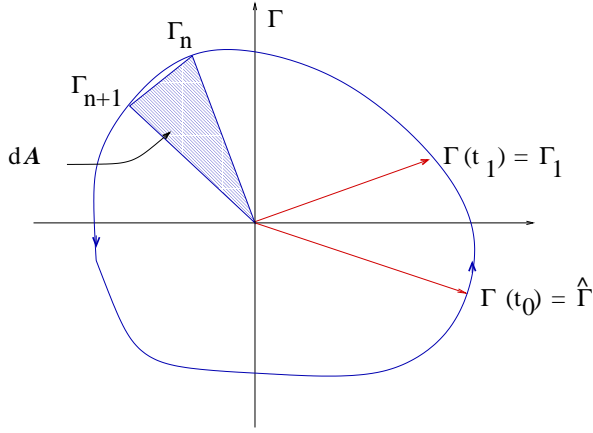}
\caption{Path of the impurity coupling $\Gamma$ in the complex plane}
\label{path}
\end{figure}

The approach introduced above can be used to find the current at any
other time by using the following scheme. We divide the entire path
traversed by $\Gamma$ into a sequence of $N$ sub-intervals, as shown
in Fig.~{\ref{path}}, starting from the equilibrium state at time
$t=t_{0}$ with a coupling $\Gamma(t_0)=\hat\Gamma$, and returning to
the same value at $t_N=T$.  The length of the $n$-th interval
$[t_{n},t_{n+1}]$ is chosen such that (for $t\in [t_{n},t_{n+1}]$) the
phase change $\delta\phi (t)=\phi(t)-\phi(t_n)$ is small enough for
the linear approximation to hold. The current $I(t)$ for $t\in
[t_{n},t_{n+1}]$ is evaluated by going to the interaction
representation with respect to the Hamiltonian ${\cal H}[\Gamma_n]={H} +
H_{\Gamma_n}$. Here $H_{\Gamma_n}$ has a time-dependent coupling
$\Gamma_n(t)$ that has a constant phase in the interval $t\in
[t_{n},t_{n+1}]$ . We can therefore write the current as:
\begin{equation}\label{ip-tau-n}
\begin{split}
  I(t)&=\left\langle S^{\dg}(t,t_n)J_{\Gamma}(t)
    S(t,t_n)\right\rangle,\\
  &{\rm where}\\
  S(t,t_n) &=T e^{-i\int_{t_n}^{t}dt'\;\delta H_{\Gamma}(t')}.
\end{split}
\end{equation}
Here $\delta H_{\Gamma}(t)$ vanishes for $t\neq [t_n,t_{n+1}]$.
Within this linear response approximation we can write the expression
for the total charge pumped in a cycle as a sum of contributions
arising from each of the intervals:
\begin{equation}
\label{qtotal}
Q =\sum_{n=0}^{N}\int_{t_{n}}^{t_{n+1}}dt\;I(t)=\sum_{n=0}^{N}\delta Q(t_{n+1})
\end{equation}
Because the various time intervals differ only in the value of the
parameters that characterize the initial time of each
interval, it is sufficient to calculate the charge pumped in any one
time interval. Contributions from all other intervals can be obtained
by appropriate relabeling of these parameters.  Thus, for
example, the charge pumped in a time interval $[t_0,t_1]$ is obtained
from (\ref{ip0}), after exchanging the order of $t$ and $t'$
integrations:
%
\begin{equation}
\label{dq1}
\begin{split}
  \delta Q(t_1)=\int_{t_0}^{t_1}dt'\;\delta\dot\phi(t')
  \int_{t'}^{t_1}dt \int_{-\infty}^{t'}dt''\; {\cal K}^{(R)}_{0}(t ;  t'')
\end{split}
\end{equation}
%
where the retarded backscattering current-current correlator for
$t\in[t_0,t_1]$ is:
\begin{equation}\label{kret}
{\cal K}^{(R)}_{0}(t ; t'')=i\theta(t-t'')
\Big\langle[J_{\Gamma_0}(t),J_{\Gamma_0}(t'')]\Big\rangle.
\end{equation}
The charge pumped in any other interval can now be found by
substituting $\Gamma_0\leftrightarrow\Gamma_n$, and
$\delta\Gamma(t)\leftrightarrow\delta\Gamma_n(t)$.  Therefore the
total charge pumped in a cycle can be expressed as:
\begin{equation}
\label{qtotal-1}
\begin{split}
  Q=\sum_{n=0}^{N}\!  \int_{t_{n-1}}^{t_n} dt'\;\delta\dot\phi (t')
  \int_{t'}^{t_n} dt\int_{-\infty}^{t'}dt''\;{\cal K}^{(R)}_{n}(t ;
  t'').
\end{split}
\end{equation}
The function multiplying $\delta\dot\phi(t')$ is to be thought of as a
generalized non-equilibrium conductance ${\cal G}$ for the
backscattering current. We write the total charge pumped in a cycle
as:
\begin{eqnarray}
  Q&=&\sum_{n=0}^{N}\!\int_{t_{n-1}}^{t_n} dt'\;
  \frac{1}{2\pi}\delta\dot\phi (t'){\cal G}_n(t'),\\
{\cal G}_n&=&2\pi\int_{t'}^{\infty}
F_n(t-t') dt\int_{-\infty}^{t'}dt''\;{\cal K}^{(R)}_{n}(t ; t''),
\end{eqnarray}
where $F_n(t-t')$ has been introduced as a smooth cut-off function
that vanishes as $t\to t_n$. In general, the quantity ${\cal G}_n$ is
dependent on the precise form of $F_n$. However, when the contribution
from the end points is negligible compared to the contribution from
the rest of the time interval $t\in(t_n,t_{n-1})$, we can expect some
interval-independent behavior. We therefore introduce a simple form
for the cut-off function, and write:
\begin{eqnarray}
\label{gn}
 {\cal G}_n
=2\pi\int_{t'}^{\infty} e^{-\omega_n (t-t')} dt\int_{-\infty}^{t'}dt''\;{\cal K}^{(R)}_{n}(t ; t''),
\end{eqnarray}
where $\omega_n\to 0^+$ in the aforementioned limit.  Next, we note
that the retarded correlator ${\cal K}^{(R)}_n$ is in general not time
translation invariant, because of the time-dependent coupling
$\Gamma(t)$.  The function $\int_{-\infty}^{t'}dt''\;{\cal
  K}^{(R)}_{n}(t ; t'')$ has a maximum along $t'=t$. For
non-interacting electrons, because of the singular nature of
fermionic Green function, the retarded correlator ${\cal
  K}^{(R)}_n(t;t'')$ has a maximum near $t''\approx t + 1/\Lambda_0$,
where $\Lambda_0$ is the upper cut-off of the problem, {\it i.e.},
$\Lambda_0\leq E_F$ (see Section~\ref{discussion} for details).
Also the width of this maximum $\delta t^*\sim1/\Lambda_0$. As a
result, for a frequency $\Omega\ll\Lambda_0$ of variation of
$\Gamma$, we can treat the time-dependent parameters
$\Gamma(t'')=\Gamma(t)=\Gamma(t')$, with an error of the order of
$\Omega/\Lambda_0\ll 1$. Keeping in mind the other energy scale in
the problem (the inverse length scale of the barrier,
$\Lambda_\ell$, which defines the validity of single impurity
approximation), the {\it adiabatic limit} is $\Omega\ll{\rm
min}[\Lambda_0,\Lambda_\ell]$. In this limit we can replace the
time-dependent Hamiltonian ${\cal H}[\Gamma(t)]$ by the {\it
instantaneous} (or static) Hamiltonian ${\cal H}[\Gamma(t')]$, and
treat the retarded current correlator as time translation
invariant.

In the presence of electron-electron interactions in the bulk, the
nature of singularity of the fermionic Green function is modified.
This deformation of the Fermi liquid picture implies that a new energy
scale enters into the problem, thereby altering the aforementioned
adiabatic criterion. In order to identify the correct adiabatic limit,
we define dimensionless integration variables: $\bar t=\Lambda
t;\,\,\bar t''=\Lambda t''$, whereby the integrand acquires a
multiplicative factor of $\Lambda^{-2}$, and rewrite ${\cal G}_n$ as:
\begin{equation}
  \label{i-scaled}
 {\cal G}_n=2\pi\int_{0}^{\infty}\!\!\!e^{-\frac{\omega_n}{\Lambda}\bar t}
 d\bar t\int_{-\infty}^{0}d\bar t''\; \Lambda^{-2}
{\cal K}^{(R)}_{n}(\frac{\bar t}{\Lambda} + t' ; \frac{\bar t''}{\Lambda}+ t').
\end{equation}
Because of interactions in the wire, the composite operator acquires
anomalous dimensions. We write the scaling dimension $\Delta$ for the
composite operator ${\cal Q}$ as $\Delta=1-a$, where $a$ is the
anomalous dimension, and invoke the scaling hypothesis (for the
equilibrium problem scaling holds whenever $|\Gamma|^2\ll 1$ for the
range $0<\Lambda\le\Lambda_0$; in the non-equilibrium case it is
further restricted to $\Omega\lesssim\Lambda\leq\Lambda_0$). Since the actual
dimension of ${\cal K}$ is fixed, we obtain:
\begin{equation}
  \begin{split}
{\cal K}^{(R)}_{n}&(\bar t/\Lambda + t' ; \bar t''/\Lambda+ t')\\
&=\Lambda^{2}(\Lambda_0/\Lambda)^{2a}{\cal K}^{(R)}_{n}(\bar t+\Lambda
t' ; \bar t''+\Lambda t').
\end{split}
\end{equation}
As a consequence, the dimensionless coupling constant
$\Gamma\to\tilde\Gamma\equiv (\Lambda_0/\Lambda)^a\Gamma$, and its
time-dependence takes the form: $\Gamma(\Omega\bar t/\Lambda+\Omega
t')$.  We can therefore identify an energy scale
$\omega_\Gamma=\Lambda\left\vert\tilde\Gamma\right\vert^{1/a}$,
dependent only on the bare coupling $\Gamma$ and the upper cut-off
$\Lambda_0$. Choosing $\Omega\lesssim\Lambda\ll\Lambda_0$ ensures
that, for $a>0$, the renormalized (dimensionless) coupling constant
$\tilde\Gamma\gg\Gamma$.~\footnote{Besides the local backscattering
  operator ${\cal Q}$ that we explicitly consider, there are other
  operators present in the bare Hamiltonian, that correspond to
  multi-particle backscattering processes~\cite{kane92b}. In the bare
  Hamiltonian each of these operators, ${\cal Q}'$, has a different
  scaling dimension, $\Delta'$, and a different (dimensionless) bare
  coupling constant, $\Gamma'$, which involves
  $2nk_F$-Fourier-transform ($n>1$) of the potential $V(x)$. As should
  be clear from the analysis presented here, it is only the most
  relevant operator that determines the conductance in the adiabatic
  limit.} The corresponding energy scale $\omega_\Gamma$ defines a
crossover scale that separates two qualitatively different responses
to the time-dependence: (i) For $\Omega\ll\omega_\Gamma$, the response
is an {\it adiabatic} modification of the ground state with a
renormalized coupling $\tilde\Gamma$, and (ii) for
$\Omega\gg\omega_\Gamma$, the response is a {\it sudden} modification
of the ground state with $\Gamma=0$. In other words, the singular
contribution from the retarded correlator ${\cal K}^{(R)}_n(t)$
acquires a width of the order of $1/\omega_\Gamma$, as compared to
$1/\Lambda_0$ for the non-interacting case.  For example, in
Section~\ref{sec-khalf} we show that, for the Luttinger model with
interaction parameter $K=1/2$, the singularity of ${\cal
  K}^{(R)}_n(t)$ at $t\sim1/\Lambda_0$ has a width $\delta t^*\sim
1/(\Lambda_0|\Gamma|^2)$.  Thus, if $\Lambda/\omega_\Gamma\ll 1$,
then we can treat $\Gamma(t'')\approx\Gamma(t)\approx\Gamma(t')$.
 Therefore, ${\cal G}_n$ can be approximated by:
\begin{equation}
  \label{i-rg}
 {\cal G}_n(t') \simeq 2\pi
\int_{0}^{\infty}e^{-\omega_n t}\; d t\int_{-\infty}^{0}d t''\;
\tilde{\cal K}^{(R)}_n(t - t''),
\end{equation}
where $\omega_n\to 0$, because the intervals $[t_{n-1},t_n]$ have lengths
$\sim 1/\Lambda\gg 1/\omega_\Gamma$; $\tilde{\cal K}^{(R)}_n(t-t'')$ is the
retarded backscattering current correlator with the renormalized
coupling constant $\tilde\Gamma(t')$, independent of the integration
variables $t,t''$. It is important to note that the retarded
correlator $\tilde{\cal K}$ is time-translation invariant in the
arguments $\bar t,\,\bar t''$. This transformation, from a current
correlator without time translation invariance to a renormalized
correlator with time translation invariance, is explicitly shown in
Section~\ref{k-half-sol} for the special cases when the scaling
dimension of ${\cal Q}$ is $1$, and $1/2$. A consequence of this time
translation invariance is that we can rewrite the time integrals to
obtain:
\begin{equation}
  \label{irg-g}
\begin{split}
  {\cal G}_n\simeq2\pi\int_{0}^{\infty}d\tau \left[\frac{1-e^{-\omega_n
        \tau}}{\omega_n}\right] \; \tilde {\cal K}^{(R)}_n(\tau)
\end{split}
\end{equation}
The integral involving $\tilde{\cal K}^{(R)}_n$ can be readily
identified as the conductance for backscattering current at frequency
$i\omega_n$.
\begin{equation}
{\cal G}_{\tilde\Gamma(t')}=2\pi\frac{\tilde {\cal
K}^{(R)}_n(0+i\omega_n)-\tilde {\cal K}^{(R)}_n(0)}{\omega_n}.
\label{gb}
\end{equation}
Here $\tilde{\cal K}^{(R)}_n(0+i\omega_n)$ represents the Fourier
transform of $\tilde {\cal K}^{(R)}_n(t)$ at frequency $\omega_n$
analytically continued to imaginary frequency $i\omega_n$ (with
$\omega_n\to 0^+$). We emphasize that the conductance depends on the
renormalized coupling $\tilde\Gamma(t')$ which is time-dependent. As a
consequence we can write the total charge pumped in a cycle as an
integral over the path of this coupling constant:
\begin{equation}\label{qb-cond}
\begin{split}
Q&=\int_{\phi(t')=\phi_0}^{\phi(t')=\phi_0+2\pi} dt'\;\dot\phi(t')\;
{\cal G}_{\tilde\Gamma(t')}\\
&=\frac{1}{2\pi i}\oint \frac{d\Gamma}{\Gamma}{\cal G}_{\tilde\Gamma}.
\end{split}
\end{equation}
%
%
Thus the problem of evaluating $Q$, for $\Omega\ll\omega_\Gamma$,
reduces to calculating the dependence of the backscattering
conductance on the renormalized impurity couplings.  In the next
section we shall link this quantity to the more familiar conductance
of the fermionic current.

\subsection{Pumped charge and dc conductance}
\label{dccond}
%

We now proceed to recast the expression (\ref{qb-cond}) in terms of the response
function for fermionic currents $\tilde j$ to a vector potential $A(t)$:
\begin{equation}
\begin{split}
G=\frac{\delta\langle\tilde j(x,t)\rangle}{\delta A(t')}.
\end{split}
\end{equation}
Let us recall that in the case of linear response to a perturbative
$A$ which vanishes outside a region of length $L$, the conductance is
given by (see, for example, Ref~\onlinecite{kane92b})
\begin{equation}
\begin{split}
G(\omega)=\frac{1}{L\omega}\int_{-L/2}^{L/2}\!\! dy
\int_{-\infty}^{0}\!\! dt\;(1-e^{-i\omega t})\;\langle[\tilde
j(0,0),\tilde j(y,t)]\rangle.
\label{gcond}
\end{split}
\end{equation}

In order to find an expression for pumped current in the low frequency
($\Omega\lesssim\omega_\Gamma$) limit, we begin by writing
the fermionic current at any point $x$ in the wire at time
$t\in[t_{n},t_{n+1}]$, in a way similar to (\ref{ip-t}) of
Section~\ref{chargepump}, albeit now with the effective action
with renormalized couplings $\tilde\Gamma$:
\begin{equation}\label{ipx-t}
I(x,t)=\left\langle S^{\dg}(t,t_{n})\,\tilde
j(x,t)\,S(t,t_{n})\right\rangle,
\end{equation}
Expanding this perturbatively to first order in the phase change
$\delta\phi(t)$ of the coupling $\Gamma(t)$ in the time interval
$[t_{n},t_{n+1}]$, and using the notations of Section~\ref{chargepump}
we get:
\begin{equation}\label{delta-ipx}
\begin{split}
I(x,t)& = \left\langle \tilde j(x,t)\right\rangle\\ &- i
\int_{t_{n}}^{t}dt'\delta\phi(t') \left\langle[\tilde j(x,t),
J_{\tilde\Gamma_n}(t')]\right\rangle.
\end{split}
\end{equation}
We can now invoke Eq.~(\ref{jgamma}) and do an integration by parts to
write the current as:
\begin{equation}\label{ipx0}
\begin{split}
  I(x,t) &= \left\langle \tilde j(x,t)\right\rangle\\ &-
  i\int_{t_n}^{t}\!
  dt'\;\delta\dot\phi(t')\int_{-1/\epsilon}^{1/\epsilon}
  dy\left\langle[\tilde j(x,t), \tilde j(y,t')]\right\rangle,
\end{split}
\end{equation}
where $\epsilon\equiv 2/L \lesssim \Omega$.
Since the unperturbed Hamiltonian, under which all operators evolve,
involves $\tilde\Gamma_n(t)$ which has a constant phase in the time
interval $[t_n,t_{n+1}]$, it follows that $\langle\tilde
j(x,t)\rangle=0$.  The pumped charge in the same interval can be
written as:
\begin{equation}
  \label{dqx-1}
  \delta Q(t_n)=
  -\int_{t_n}^{t_{n+1}}dt'\delta\dot\phi(t')\int_{t'}^{t_{n+1}}\!
 dt\; K^{(R)}_n(t,t')
\end{equation}
where the retarded current correlator:
$$
K^{(R)}_n(t,t')=i\int_{-1/\epsilon}^{1/\epsilon}
dy\;\theta(t-t')\;\Big\langle[\tilde j(x,t),\tilde j(y,t')
]\Big\rangle,
$$
and the subscript $n$ reminds us that this depends on the time
dependent coupling $|\tilde\Gamma_n(t)|$.  In the slow pumping limit the contribution to the integral
over $t$ from the end-point $t_{n+1}$ is negligible (or the integral
is dominated by the singular contribution near $t\approx t'$),
expression (\ref{dqx-1}) can be rewritten as:
\begin{subequations}
\label{dqx1}
\begin{eqnarray}
\label{dqx1-a}
\delta Q(t_n)
&=&-\int_{t_n}^{t_{n+1}}dt'\;\delta\dot\phi(t')\int_{t'}^{\infty}\!
dt\; K^{(R)}_n(t,t')\\
+\int_{t_n}^{t_{n+1}}&dt'&\;\delta\dot\phi(t')\int_{t'}^{\infty}\!\!\!dt
\left[1-e^{-\epsilon t}\right] K^{(R)}_n(t,t').
\label{dqx1-b}
\end{eqnarray}
%
\end{subequations}
As discussed in section \ref{cpump-gb}, when
$\Omega/\omega_\Gamma\ll 1$, we can obtain the effective
Hamiltonian that evolves the operators $\tilde j$. This effective
Hamiltonian has static renormalized couplings $\tilde\Gamma_n(t')$,
yielding a time-translation invariant correlator:
$$
K_n^{(R)}(t,t')\to\tilde K_n^{(R)}(t-t').
$$
In this limit the
expression (\ref{dqx1-b}) above, can be identified as the equilibrium
dc conductance $\lim_{\epsilon\to
  0^+}G_{\tilde\Gamma_n}(0+i\epsilon)$:
\begin{equation}
\begin{split}
\label{gcondi}
G_{\tilde\Gamma_n}(0+i\epsilon)=&-\frac{1}{2}
\int_{0}^{\infty}dt \left[1-e^{-\epsilon t}\right]\\
\times &\int_{-1/|\epsilon|}^{1/|\epsilon|} dy
\;i\Big\langle[\tilde j(0^- , t), \tilde j(y,0)]\Big\rangle_{\tilde\Gamma_n(t')},
\end{split}
\end{equation}
where the subscript $n$ reminds us of the renormalized coupling
constant $\tilde\Gamma_n$. On the other hand, expression
(\ref{dqx1-a}) can be simplified using the current continuity equation
$$
\partial_{x'}\tilde j(x',t)=\partial_t\tilde \rho(x',t);\,x'\neq 0,
$$
inside the expectation value:
\begin{equation}
\begin{split}
&\int_{t'}^{\infty}dt\;
\frac{1}{2i}\int_{-1/\epsilon}^{1/\epsilon} dy\;\Big\langle[\tilde j(x,t), \tilde j(y,t')]
\Big\rangle\\
=&\int_{t'}^{\infty}dt \;
\frac{1}{2i}\int_{-1/\epsilon}^{1/\epsilon} dy\int_{-\infty}^{x=0^-}dx'
\Big\langle[\partial_{x'}\tilde j(x',t), \tilde j(y,t')]\Big\rangle\\
=&\frac{1}{2i}\int_{-1/\epsilon}^{1/\epsilon} dy \int_{-\infty}^{x=0^-}dx'
\int_{t'}^{\infty}dt \;\partial_t \Big\langle[\rho(x',t), \tilde j(y,t')]\Big\rangle\\
=&\frac{i}{2}\int_{-1/\epsilon}^{1/\epsilon} dy \int_{-\infty}^{x=0^-}dx'
\Big\langle[\rho(x',t'), \tilde j(y,t')]\Big\rangle\\
=&\frac{i}{2}\int_{-1/\epsilon}^{1/\epsilon} dy \int_{-\infty}^{x=0^-}dx'
\left(\frac{-i}{\pi}\right)\partial_{x'}\delta(x'-y)=\frac{1}{2\pi}.
\end{split}
\end{equation}
Here, in the last step, we have used the equal time anomalous
commutator of fermionic density and current,
\begin{equation}
\label{an-com}
[\rho(x,t'),\tilde j(y,t')]=-\frac{i}{\pi}\partial_x\delta(x-y).
\end{equation}

Putting all the terms together we get:
%
\begin{equation}
\label{dqx1-g}
\delta Q(t_n)=
\int_{t_n}^{t_{n+1}}dt'\;\delta\dot\phi(t')
\Big[\frac{1}{2\pi} - G_{\tilde\Gamma_n}(0+i\epsilon) \Big]
\end{equation}
%
The prescription to obtain the total charge pumped in a cycle, $Q$, is
exactly the same as in Section~\ref{cpump-gb}: divide the entire path
of $\Gamma(t)$ into appropriate intervals, and find the charge pumped
in each interval in the linear response approximation.  We thus find
the charge pumped through $x=0$ in the limit $\Omega\ll\omega_\Gamma$:
\begin{equation}\label{qb-dccond}
\begin{split}
Q=\frac{1}{2\pi i}\oint \frac{d\Gamma}{\Gamma} \left[1-
\lim_{\epsilon\to 0}2\pi G_{\tilde\Gamma}(0+i\epsilon)\right].
\end{split}
\end{equation}
%
%


\section{Transport of charge and spin using parametric pumping}
\label{spin-trans}
\begin{figure}
\includegraphics[scale=.45]{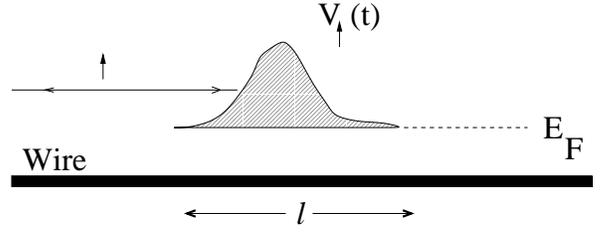}
\caption{Scattering potentials for up ($\up$) and down ($\dwn$) spins whose variation in time drives the spin pump}
\label{voex}
\end{figure}
We now include spin to analyze both the {\it spin pump}
and the {\it charge pump} of Ref.~\onlinecite{sharma01}. We begin by
considering a Hamiltonian $H$ which conserves the $z$ component of the
total spin, and chirality in each of the two spin states ($\up,\dwn$). Thus
the axial charges
\begin{equation}
\int dx \;\tilde j^{c,s}(x,t)=\int dx \left[\tilde j^\up(x,t) \pm
\tilde j^\dwn(x,t)\right]
\end{equation}
are conserved. Here $\tilde j^{c,s}$ are the fermion current operators
for charge and spin currents.  The spin (charge) pump operates by
adding to $H$ a time-dependent potential, locally breaking the chiral
symmetry in the spin (charge) sector.  For the spin pump we consider a
localized magnetic field in the $z$ direction, acting as an impurity
potential $V_o(x,t)$ that couples anti-symmetrically to up ($\up$) and
down ($\dwn$) spin states. In addition we consider a localized
scattering potential $V_e(x,t)$, coupling symmetrically to the spin
states. This is shown schematically in Figure \ref{voex}, where we
introduce the notation $V_{\up,\dwn}\equiv V_e\pm V_o$.
\begin{figure}
\includegraphics[scale=.35]{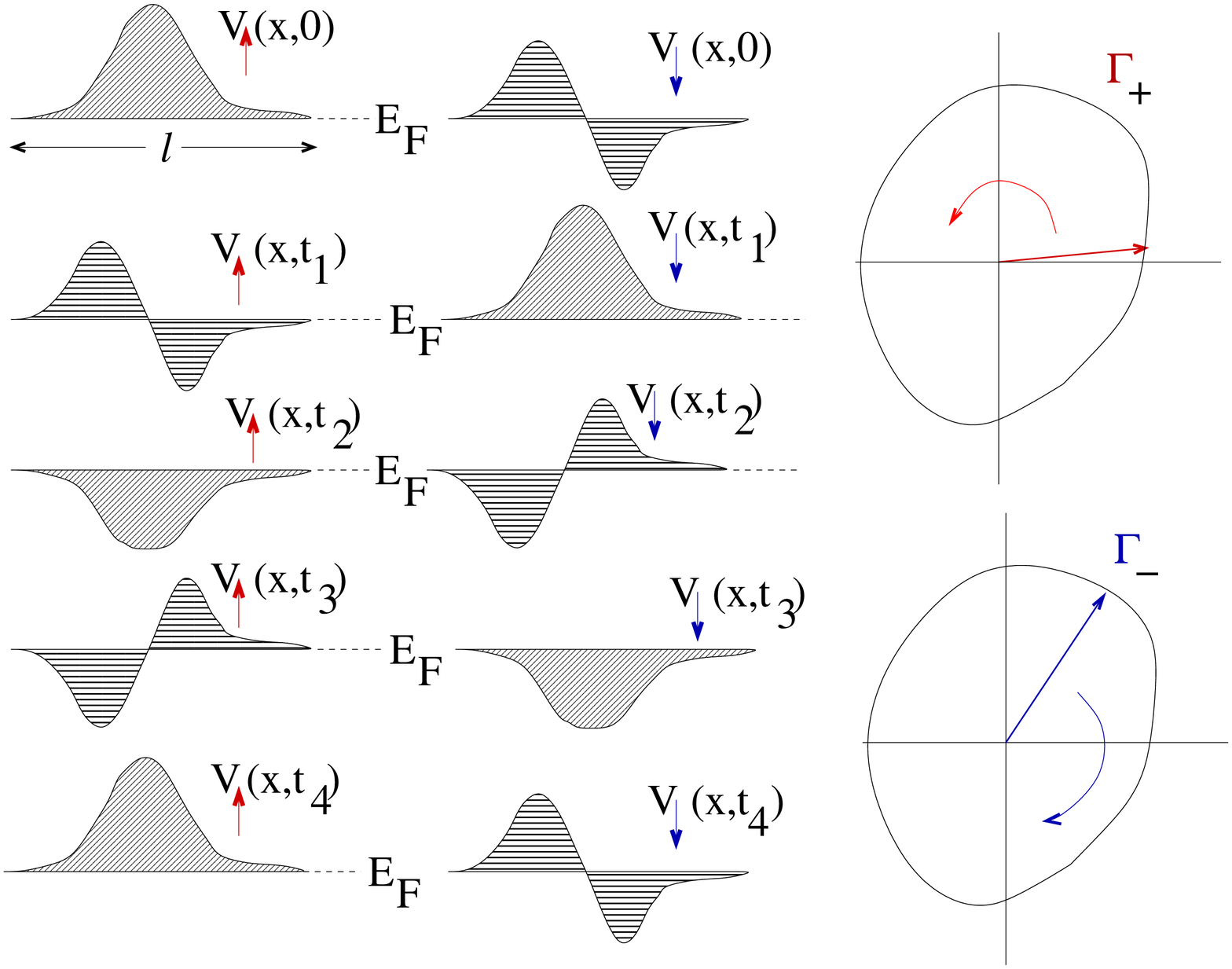}
\caption{Scattering potentials for up ($\up$) and down ($\dwn$) spins
whose variation in time gives rise to opposite sense of rotation for
$\Gamma_\pm$}
\label{gammapm}
\end{figure}
The additional term in the Hamiltonian is:
\begin{equation}\label{voe-sc}
\begin{split}
  \int_{-l}^{l} dx\; V_{e}(x)\left[\psi^{\dg}_{\up}(x)\psi_{\up}(x)+\psi^{\dg}_{\dwn}(x)\psi_{\dwn}(x)\right]\\
  +\int_{-l}^{l} dx\;
  V_{o}(x)\left[\psi^{\dg}_{\up}(x)\psi_{\up}(x)-\psi^{\dg}_{\dwn}(x)\psi_{\dwn}(x)\right]
\end{split}
\end{equation}
As we pointed out before in Section \ref{cp-model}, it is the backward
scattering of the right and left movers from the potentials that leads
to a non-zero current.  Following Ref.~\onlinecite{sharma01} we
consider the case when the potentials
$V_{o,e}(x,t)=V_{o,e}(x)f_{o,e}(t)$ with real periodic functions
$f_{o,e}(t)$. In particular, for harmonic pumping we can choose
$f_e(t)=\cos(\Omega t+\theta)$ and $f_o(t)=\cos\Omega t$. The elastic
backward scattering of spin up (down) electrons is then given by the
complex quantities:
\begin{equation}
  \label{gm-pm}
\begin{split}
  \Gamma_\pm(t)
  &=\int dx\; e^{i2k_Fx}\;\left[V_e(x)f_e(t)\pm V_o(x)f_o(t)\right]\\
  &=\tilde V_e(2k_F)f_e(t)\pm\tilde V_o(2k_F)f_o(t)
\end{split}
\end{equation}
which are the $2k_F$-Fourier- transform of the potentials that scatter
up or down ($\pm$) spin particles. Thus, if $V_e(x)$ and $V_o(x)$ are
different functions of position (both being assymetric with respect to
$x=0$), then the complex quantities $\tilde V_{e,o}(2k_F)$ have
different arguments ({\it cf.} Section~\ref{cp-model}), and by
suitably varying the real functions $f_{e,o}(t)$ (making them change
sign) we can trace any path in the complex plane which winds around
the origin.  Furthermore, the the difference between $\Gamma_{+}$ and
$\Gamma_{-}$ is that they wind in opposite directions on the complex
plane, as shown in Fig.~\ref{gammapm}.

We can denote the composite operators, which have couplings
$\Gamma_{\pm}$, by ${\cal Q}_{\pm}$. It is more convenient, however,
to define operators which transform simply under rotations in spin
space.  We therefore write the backscattering term in the Hamiltonian
(\ref{voe-sc}) by defining linear combinations of $\Gamma_\pm$:
\begin{equation}
H_{\Gamma} = \sum_{\beta=o,e}\Gamma_{\beta} {\cal Q}_{\beta}
+{\Gamma_\beta}^{*} {\cal Q}_{\beta}^{\dg},
\end{equation}
where the ${\cal Q}_{o (e)}$ field is odd (even) under the unitary
operator ${\cal R}_y(\pi)=e^{-i\frac{\pi}{2}\sigma_y}$, which does a
$\pi$-rotation about the $y$-axis in the spin space, and
$\Gamma_{e,o}=\frac{1}{2}[\Gamma_+\pm\Gamma_-]$.
The backscattering current that arises from non-conservation of the axial charges
$\int dx\; \tilde j^{c,s}(x)$ is now
\begin{equation}\label{jgsc}
\begin{split}
J^{s(c)}_{\Gamma} &=-\frac{1}{2}\int dx\; \partial_t \;\tilde j^{s(c)}(x)\\
& = i\Gamma_{o} {\cal Q}_{e(o)}+
i\Gamma_{e} {\cal Q}_{o(e)} + {\rm H.c.}.
\end{split}
\end{equation}
WT identities, similar to Eq.~(\ref{wi-1},\ref{wi-2}) can now be
derived for each of ${\cal Q}_{o,e}$ fields and their corresponding
backscattering currents. As before (Sections~\ref{cpump-gb} and
\ref{dccond}), WT identities ensure that there is no current without a
change in the phase of $\Gamma_\pm$.


\subsection{Spin pump}

We now consider the currents generated by a small change in the phase
of the couplings $\Gamma_{\pm}=\Gamma_e\pm\Gamma_o$, in a time
interval $t\in [t_0,t_1]$:
%
\begin{equation}
\begin{split}
  I^{s(c)}(x,t)=-\left\langle \tilde j^{s(c)}(x,t)\right\rangle
  +\sum_{\sigma=\pm} \int_{0}^{t} dt'\;\delta\phi_{\sigma}(t')\\
  \times i\left\langle[\tilde j^{s(c)}(x,t),
    J^{c}_{\tilde\Gamma_0}(t')+\sigma J^s_{\tilde\Gamma_0}(t')]\right\rangle,
\end{split}
\end{equation}
%
where, as in Section~\ref{dccond}, $\tilde\Gamma_0(t)$ is the time
dependent renormalized coupling with a constant phase for
$t\in[t_0,t_1]$.  Next, we invoke Eq.~(\ref{jgsc}), albeit with
renormalized couplings, and do an integration by parts to write the
current as:
\begin{equation}
\begin{split}
  I^{s(c)}(x,t)&=\sum_{\sigma=\pm}\int_{t_0}^{t} dt'
  \frac{1}{2}\delta\dot\phi_{\sigma}(t')\int_{-\infty}^{t'}dt''\\
  &\times\int_{-1/\epsilon}^{1/\epsilon} dy\;i\left\langle[\tilde
    j^{s(c)}(x,t), \tilde j^{c}(y,t'')+\sigma\tilde j^s (y,t'')]
  \right\rangle,
\end{split}
\end{equation}
%
where $\epsilon=2/L\lesssim\Omega$.
The pumped spin (charge) in this interval can now be written as:
\begin{equation}
\label{dqscx-1}
\begin{split}
  \delta Q^{s(c)}(t_0)
  &=\int_{t_0}^{t_{1}}dt'\sum_{\sigma=\pm}\delta\dot\phi_\sigma (t')\\
  &\times\int_{t'}^{t_{1}}\! dt\;\frac{1}{2}\left[
    K^{(R)}_{0;s(c),c}(t,t')+ \sigma K^{(R)}_{0;s(c),s}(t,t')\right]
\end{split}
\end{equation}
where the retarded current correlator:
$$
K^{(R)}_{0;a,b}(t,t')=i\int_{-1/\epsilon}^{1/\epsilon}
dy\;\theta(t-t')\;\Big\langle[\tilde j^a(x,t),\tilde j^b(y,t')
]\Big\rangle.
$$
Following Section \ref{dccond}, we rewrite this expression as:
\begin{equation}
\label{dqscx-2}
\begin{split}
  \delta Q^{s(c)}(t_0)
  &=\int_{t_0}^{t_{1}}dt'\sum_{\sigma=\pm}\delta\dot\phi_\sigma (t')\\
  &\times\int_{0}^{\infty}\!e^{-\epsilon t}\; dt\;\frac{1}{2}\left[
   \tilde K^{(R)}_{0;s(c),c}(t)+ \sigma
  \tilde K^{(R)}_{0;s(c),s}(t)\right],
\end{split}
\end{equation}
which can be recast as:
\begin{equation}
  \label{dqscx-3}
\begin{split}
 \delta Q^{s(c)}(t_0)&=\int_{t_0}^{t_{1}}dt'\sum_{\sigma=\pm}\delta\dot\phi_\sigma (t')\\
&\times\int_{0}^{\infty}dt\;\frac{1}{2}\left[ \tilde K^{(R)}_{0;s(c),c}(t)+\sigma \tilde K^{(R)}_{0;s(c),s}(t)\right]\\
&-\int_{t_0}^{t_{1}}dt'\sum_{\sigma=\pm}\frac{1}{2}\delta\dot\phi_\sigma (t')\\
&\times\left[G^{s(c),c}_{\tilde\Gamma}(0+i\epsilon)+\sigma G^{s(c),s}_{\tilde\Gamma}(0+i\epsilon)\right],
\end{split}
\end{equation}
where
\begin{equation}
\begin{split}
\label{gsccondi}
G^{a,b}_{\tilde\Gamma}(0+i\epsilon)=&-\frac{1}{2}
\int_{0}^{\infty}dt \left[1-e^{-\epsilon t}\right]\\
\times &\int_{-1/|\epsilon|}^{1/|\epsilon|} dy
\;i\Big\langle[\tilde j^a(0^- , t), \tilde j^b(y,0)]\Big\rangle.
\end{split}
\end{equation}
The mixed conductance $G^{s,c}_{\tilde\Gamma}=0$, from the following
symmetry argument: The Hamiltonian (with the impurity) is invariant
under a combination of spin rotation ${\cal R}_y(\pi)$ and local
magnetic field inversion, whereas the conductance
$G^{s,c}_{\tilde\Gamma}\to - G^{s,c}_{\tilde\Gamma}$.  We can now use
a less loaded notation by labeling the spin (charge) conductance by a
single superscript.
Thus we obtain the following expression for total spin and charge transported through any point in the
wire in a pumping cycle:
%
\begin{equation}
\begin{split}
\label{qsc}
Q^{s}=\sum_{\sigma=\pm}\sigma\frac{1}{2\pi i}\oint\frac{d\Gamma_{\sigma}}{\Gamma_{\sigma}}
\left[1-\pi G^{s}_{\tilde\Gamma}\right]  \\
Q^{c}=\sum_{\sigma=\pm}\frac{1}{2\pi i}\oint\frac{d\Gamma_{\sigma}}{\Gamma_{\sigma}}
\left[1-\pi G^{c}_{\tilde\Gamma}\right].
\end{split}
\end{equation}
Here $G^{c(s)}_{\tilde\Gamma}$ is the DC charge (spin) conductance with
renormalized impurity coupling constants $\tilde\Gamma_{\pm}$.  For the
{\it spin  pump} the choice of time dependence for the parameters
$V_{o,e}(x,t)=V_{o,e}(x)f_{o,e}(t)$ (discussed earlier) implies that
\begin{equation}
\oint\frac{d\Gamma_{+}}{\Gamma_{+}}=-\oint\frac{d\Gamma_{-}}{\Gamma_{-}}=2\pi i.
\end{equation}
We can now write the pumped spin (in integer values of $\hbar/2$) as:
\begin{eqnarray}\label{spump-qsc}
Q^s = 2 -\sum_{\sigma=\pm}\frac{1}{2\pi i}\oint \frac{d\Gamma_{\sigma}}{\Gamma_{\sigma}}
\sigma\pi G^{s}_{\tilde\Gamma}\\
Q^c=-\sum_{\sigma=\pm}\frac{1}{2\pi i}\oint \frac{d\Gamma_{\sigma}}{\Gamma_{\sigma}}
\pi G^{c}_{\tilde\Gamma}.
 \end{eqnarray}

\subsection{Charge pump}
The charge pump operates by replacing the magnetic field ($V_o(x,t)$)
by a charge (spin independent) scattering potential. The above
analysis, and especially the expressions (\ref{qsc}), can be
straightforwardly adapted to this case by choosing
$\Gamma_+(t)=\Gamma_-(t)$.  We get the charge pumped (in units of $e$)
in a cycle:
\begin{eqnarray}\label{cpump-qc}
Q^c = 2 -\frac{1}{2\pi i}\oint \frac{d\Gamma_{+}}{\Gamma_{+}}2\pi G^{c}_{\tilde\Gamma},
 \end{eqnarray}
while no spin is pumped.

\section{Discussion: Probing the
  fixed points of the one dimensional impurity problem}
\label{discussion}

In this section we discuss the behavior of pumped charge and spin for
both the spin and charge pumps discussed in earlier sections. We find
that bulk interactions play a crucial role in determining what the
pumped charge or spin will be in both the asymptotic limits of low
($\Omega\ll \omega_\Gamma$) and high ($\Omega\gg \omega_\Gamma$)
frequency temporal variations, where the crossover energy scale
$\omega_\Gamma$ has been introduced earlier. By contrast, for
non-interacting electrons there is no such distinction. As we discuss
below, it is the non-Fermi liquid behavior peculiar to
one-dimensional interacting fermion systems, which is responsible for
this difference. As a result we find that the pumped charge or spin
behavior at low pumping frequency probes the fixed point behavior of
the equilibrium impurity problem.

The low frequency behavior of pumped charge and spin can be obtained
from the expressions (\ref{qb-cond}), (\ref{qb-dccond}),
(\ref{spump-qsc}) and(\ref{cpump-qc}), derived earlier. We first
consider the case of non-interacting electrons.

\subsection{Non-interacting electrons}
\label{lowfreq-ni}

In Section \ref{cp-model} we had pointed out that the impurity
coupling $\Gamma$ has zero dimension. In the absence of interactions
this is also the scaling dimension of the renormalized coupling
$\tilde\Gamma$, which is thereby {\it marginal} in the RG sense.  This
absence of scaling with a change of cut-off (in the notation of
Section~\ref{cpump-gb}, the exponent $a=0$) means that if the barrier
$V(x)\ll E_F$, one can evaluate the conductance (of both the fermionic
current and the backscattering current) at any energy scale using
perturbation theory.  To lowest order in the coupling $\Gamma$ (a more
complete calculation is given in Section~\ref{keq-one}) we proceed as
follows.

\subsubsection{Free electron Propagator and the Current Correlator}

In order to calculate the retarded backscattering current-current
correlator ${\cal K}^{(R)}(t,t'')$ perturbatively in the
time-dependent coupling $\Gamma$, we need the (local in space) free
electron propagator for each of the left (right) moving fermions. With
our normalization (a constant density of states per unit length $\nu_0=1/\pi$, and a
cut-off $\Lambda_0=E_F$) this is:
\begin{equation}
  D(t)=\frac{\nu_0}{it+{\rm sgn}(t)/\Lambda_0}.
\end{equation}
The zeroth order term of ${\cal K}^{(R)}(t,t'')$ is then given by:
\begin{equation}
  {\cal K}^{(R)}(t,t'')=
|\Gamma(t)||\Gamma(t'')|
\frac{4(t-t'')}{\Lambda_0}\left[\frac{\nu_0}{(t-t'')^2+1/\Lambda^2_0}\right]^2.
\end{equation}
If the frequency of variation of $|\Gamma(t'')|$ is much smaller
than $1/\Lambda_0$, then we can approximate $|\Gamma(t'')|\approx
|\Gamma(t)|$ inside the integral $\int_{-\infty}^{t'}dt''{\cal
  K}^{(R)}(t;t'')$, as it is dominated by the contribution near
$t''\approx t-1/\Lambda_0$. Furthermore, we can write
$|\Gamma(t)|\approx |\Gamma(t')|$. These approximations can be
performed at every order in the expansion of ${\cal K}^{(R)}$, so
that we obtain, from
$\int_{t'}^{\infty}dt\int_{-\infty}^{t'}dt''{\cal
K}^{(R)}(t,t'')$, a DC backscattering conductance which depends
only the {\it instantaneous} parameter $|\Gamma(t')|$. For the
spinless case we find the lowest order contribution to the DC
backscattering conductance:
$$
G_\Gamma \simeq \frac{1}{\pi} |\Gamma|^2,
$$
Using this expression in Eq.~(\ref{qb-cond})
we get a non-quantized charge
$$
Q\simeq \frac{1}{i\pi}\oint\left[ \Gamma^* d\Gamma-\Gamma
  d\Gamma^*\right].
$$
The integral over the path can be rewritten (using Stoke's theorem) as
an integral over the area of the contour:
$$Q\simeq \frac{1}{\pi}\int_{\cal A}d{\cal A},$$
where the
differential area element $d{\cal A}$ is shown in Fig.~\ref{path}.
The charge pumped is thus dependent only on the area covered in a
cycle on the $\Gamma$-plane, not on where the area is located in the
$\Gamma$-plane.

\subsection{Interacting electrons: Luttinger liquid behavior, spinless case}
\label{lowfreq}

In the presence of interactions in the one-dimensional system the
renormalized coupling $\tilde\Gamma$ acquires {\it anomalous}
dimension $a$, which is indicative of the absence of a quasi-particle
pole in the (bulk) electronic Green function:
$$\tilde D(k)\sim (|k|-k_F)^{-1+a};
$$
$a=0$ for non-interacting electrons where we recover the
quasi-particle pole.  As a result, the coupling
$\tilde\Gamma(\Lambda)$ changes with the cut-off $\Lambda$.  In the
language of the RG, the coupling is {\it relevant} when $a>0$ and {\it
  irrelevant} when $a<0$:
$$-\frac{d\tilde\Gamma}{d\ln\Lambda}=a\tilde\Gamma$$

It is well-known~\cite{kane92b,furusaki96} that the
backscattering impurity coupling is always {\it relevant} when the
electron-electron interactions are repulsive so that
$\tilde\Gamma(\Lambda)$ increases as $\Lambda\to\Omega$, the frequency
of pumping. As $\Omega\to 0$ the DC conductance at zero temperature
vanishes. This implies that the charge pumped in a cycle (\ref{qb-dccond})
in the asymptotic limit of slow pumping is:
\begin{equation}
\begin{split}
Q=\frac{1}{2\pi i}\oint \frac{d\Gamma}{\Gamma}
\left[1- \lim_{\epsilon\to 0}2\pi G_{\tilde\Gamma}(0+i\epsilon)\right]\\
=\frac{1}{2\pi i}\oint \frac{d\Gamma}{\Gamma}=1
\end{split}
\end{equation}

\subsubsection{Frequency corrections}

Let us now turn to the question of low-frequency corrections to the
quantization of $Q$.  As $\Omega$ approaches $\omega_\Gamma$, our
assumption of replacing the upper limit of integration over variable
$t$ in expression (\ref{gn}), or (\ref{dqx-1}), by a cut-off
$\omega_n\to 0^+$, is no longer accurate.  Deviations from this arise
as the interval length $|t_n-t_{n-1}|$ shortens with increase of
pumping frequency $\Omega$ (in order for the linear approximation to
hold). As a result, the dependence of ${\cal G}_{\tilde\Gamma_n}$ on
the end-point of the integration region, {\it i.e.}, $t_n$, in each of
the summed quantities in (\ref{qtotal-1}) grows; likewise for
$G_{\tilde\Gamma_n}$ in (\ref{dqx1-g}).  This dependence can be taken
into account very simply by choosing
$\omega_n\equiv\epsilon\sim\Omega$. Then the resulting quantity is the
conductance $G_{\tilde\Gamma}(0+i\epsilon)$, simply related, by
analytic continuation, to the real frequency conductance
$G_{\tilde\Gamma}(\epsilon)$ for $\epsilon>0$.  The functional form
$G_{\tilde\Gamma}(\epsilon)$, as $\epsilon\to 0$ is governed by the
effective impurity coupling $\tilde\Gamma(\Lambda)$ at the energy
scale $\Lambda\sim\Omega$. To determine this behavior we have to know
the finite ac-bias conductance of the interacting system with an
impurity which requires recourse to a particular model.  Exact
solutions for the Luttinger model with a time-independent impurity
coupling $\Gamma$ have been shown to have a scaling form for the
conductance $G(\omega)\equiv G([\omega/T_B])$ with the impurity
strength dependent energy scale \cite{fendley95} $T_B\sim\Lambda_0
|\Gamma|^{1/(1-K)}$, and the Luttinger parameter $K<1$ for repulsive
interactions.  In the limit $\omega\ll T_B$ the conductance vanishes
as $(\omega/T_B)^{2/K-2}$. This can be used to find the asymptotic
frequency dependence of the pumped charge as follows. We replace the
impurity energy scale $T_B$ with the instantaneous one
$\omega_\Gamma(t')$, and the energy scale set by the applied
voltage/frequency, $\omega$, by that set by the cut-off $\epsilon$.
Thus we obtain:
\begin{equation}
\begin{split}
G_{\tilde\Gamma(t')}(\epsilon)\sim\left (\frac{\epsilon}{\omega_\Gamma(t')}\right)^{2/K-2}
\end{split}
\end{equation}
The pumped charge is therefore:
\begin{equation}\label{qpump-lf}
\begin{split}
Q&\simeq 1 -\frac{1}{2\pi i}\oint \frac{d\Gamma}{\Gamma}\\
&\times\left[C_1\Omega^{2/K-2}|\Gamma|^{2/K} + C_2\Omega^{4/K-4}|\Gamma|^{4/K}+\dots\right]\\
&=1-q_1(\Omega/\bar\omega_\Gamma)^{2/K-2}-q_2(\Omega/\bar\omega_\Gamma)^{4/K-4} +\dots
\end{split}
\end{equation}
where $C_1,\,C_2$ are
cut-off dependent non-universal constants, whose ratios are universal
numbers. However, in the expansion of pumped charge $Q$ the ratios of
coefficients, for example $q_1/q_2$, as well as the quantity
$\bar\omega_\Gamma$, are dependent on the details of the path followed
by the coupling $\Gamma$ and are consequently non-universal.  We note
though, that the pumped quantity (in this case, the charge $Q$) is
bounded from above (below) by its value in the case of a constant
$|\Gamma|=\;{\rm max (min)}[|\Gamma(t)|]$, which also corresponds to
a larger (smaller) value for the energy scale $\bar\omega_\Gamma$.
The above considerations give us the frequency-dependent correction to
the quantized pumped charge $Q$ in a one-dimensional system of
spinless (spin-polarized) interacting electrons.

When the electron-electron interactions are attractive, then it is
known that the impurity coupling is {\it irrelevant}\cite{kane92b}. As
a result the renormalized impurity coupling at the energy scale
$\Lambda\sim\Omega$ is smaller than the bare coupling at energy scale
$\Lambda_0>\Lambda$ and the DC conductance $G_{\tilde\Gamma}$ can be
calculated perturbatively. The lowest order correction corresponds to
setting $\tilde\Gamma=0$ which gives zero pumped charge. For the
frequency dependent corrections we turn to the Luttinger model with
$K>1$. Using the exact results \cite{fendley95} and the scaling
arguments used earlier for the $K<1$ case, we obtain the conductance
$2\pi G_{\tilde\Gamma}\simeq 1-q'_1(\Omega/\bar\omega_\Gamma)^{2K-2}$,
and the charge pumped in a cycle vanishes with a power law: $Q\sim
(\Omega/\bar\omega_\Gamma)^{2K-2}$.

\subsection{Quantization and fractional charge}

While the above analysis for spinless electrons does not directly
apply to the case of a quantum wire -- where both spin species of
electrons are present -- it correctly describes a charge pump
operating in a quantum Hall system, where the left and right movers
are the chiral edge excitations. For integer Hall systems the chiral
excitations carry integer charge and correspond to the non-interacting
case discussed earlier. On the other hand, for fractional quantum Hall
systems whose edge excitations are described by Wen's edge-state
theory \cite{wen90,wen92}, the chiral excitations carry fractional
charge $\nu e$ and correspond (in our description of the pumping
behavior) to interacting electrons with $K\equiv\nu = 1/m<1$, where
$m$ is the odd-integer denominator which characterizes the bulk
fractional quantum Hall state. One interesting question that emerges
out of the quantization of adiabatically charge in units of the
electron charge $e$ is how charge fractionalization in strongly
correlated one-dimensional systems (including edge state tunneling in
fractional quantum Hall -- FQH -- systems) is manifest through
pumping. Consider, for example, a time-dependent backscattering
potential in a constriction of a $\nu=1/3$ FQH bar: it will lead to
the adiabatic transport of an electron charge $e$ within a (slow)
period $\tau=2\pi/\Omega$, or a pumping current $I_p =
\frac{e}{2\pi}\Omega$. To attain the same current with a
non-equilibrium voltage, one must have $I_p = \nu \frac{e^2}{h}V$
instead. So while the charge pumped in a cycle is $e$, the
relationship between the pumping frequency $\Omega$ and the
non-equilibrium voltage $V$ is $\Omega=\nu e V/\hbar$. Both the finite
bias and pumped currents derive from the phase of a time-dependent
backscattering potential, and the phase changing rates in the two
cases are related by the Josephson relation $\Omega=e^* V/\hbar$, with
$e^*=\nu e$. So even though the charged pumped in a cycle is integer,
independent measurements of the pumping frequency and voltage leading
to the same current yields the fractional charge relation between
$\Omega$ and $V$. Simply put, the existence of fractional charge in
the FQH is directly related to the quantization of the Hall
conductance \cite{Laughlin-in-Prange&Girvin}, and to the quantization
of pumped charge across the edges.

\subsection{Electrons with spin: spin and charge pump.}

The quantization of the pumped spin in a cycle, $Q^{s}$, depends on
the low energy behavior of the DC spin and charge conductances
$G^{s(c)}_{\tilde\Gamma_{o,e}}$ with renormalized impurity couplings
$\tilde\Gamma_{o,e}$. The behavior of these DC conductances can be
obtained in a manner similar to that discussed above in the context of
spinless fermions. Assuming that interactions in the one-dimensional
system respect $SU(2)$ symmetry in spin, it is known that the impurity
backscattering coupling $\tilde\Gamma_{o,e}$ is {\it relevant}
whenever the interactions are repulsive. As a consequence both the
spin and charge conductances vanish at low energies\cite{kane92b} with
power law corrections which can be determined by taking recourse to a
particular interaction model with spin.  In particular for the
Luttinger model one needs to only modify the exponents of the
renormalization factors in the expressions for $Q$ by replacing
$1/K\to 1/K_c+1/K_s$ and putting $K_s=2$ for the spin isotropic point
\cite{furusaki96,lesage97,kane92b} in order to get the behavior of
$Q^{s,c}$.  Note that the non-interacting case is given by
$K_c=2,K_s=2$, and repulsive interactions imply $K_c<2$.  For the
{\it  spin} pump in a wire described by the Luttinger model with
$K_s+K_c<2$, we obtain:
\begin{eqnarray}\label{wire-qsc}
  Q^s&\simeq& 1 -\sum_{n=1} q^{s}_n
  \left(\frac{\Omega}{\bar\omega_\Gamma}\right)^{2n(\frac{1}{K_c}+\frac{1}{K_s}-1)}
  +\dots\\
  Q^c&\simeq & \sum_{n=1}
q^{c}_n\left(\frac{\Omega}{\bar\omega_\Gamma}\right)^{2n(\frac{1}{K_c}+\frac{1}{K_s}-1)}
  +\dots
\end{eqnarray}
For the charge pump the pumped charge in a cycle is quantized just as
the spin is quantized for the spin pump, albeit with different
non-universal constants $\{q^s_n\}$, as in the case of
spinless fermions.

The more general case of pumping in a wire described by the Luttinger
model with $K_s\neq 2$ can also be studied using the general relations
derived in Equations ~(\ref{spump-qsc}) and(\ref{cpump-qc}). All we
need, to determine the pumped charge and spin in a cycle, is the
behavior of the DC conductance with renormalized impurity couplings.
When the charge (spin) DC conductance vanishes we have quantized
charge (spin) pumped in a cycle. Allowing for more general values of
$K_c$ and $K_s$ we can refer to the plots of Kane and Fisher
Ref.~[\onlinecite{kane92b}] for depicting the regions, in the
$K_c-K_s$ plane, where the $2k_F$ backscattering is relevant and where
it is irrelevant. In the former regions the spin pump will transport a
quantized spin per cycle, and the charge pump will transport a
quantized charge per cycle, whereas in the latter both will give
vanishing pumped charge.  In passing we would like to note that the
analysis of parametric pumping presented thus far does not apply to
the case when more than one backscattering operator is relevant. In
the Luttinger model this happens when $K_c<2/9,\,K_s=2$, when both the
$2k_F$ backscattering and $6k_F$ backscattering terms are relevant.


\subsection{Finite size and temperature effects}

Real experimental scenarios involve finite length systems contacted by
wide leads which are usually described by Fermi liquid theory.
This brings the length of the quantum wire $L_W$ as another important
scale in the problem.  The issue of contacts in the presence of an
impurity in a Luttinger liquid model has been dealt with extensively
in the literature \cite{furusaki96}. Here we use the relevant results.
Whenever $\Omega\gg v_F/L_W$ the (backscattering) conductance ${\cal
  G}_{\tilde\Gamma}$ is determined by the properties of the wire
alone, so that the above considerations for quantization and
corrections from non-zero $\Omega$ hold.  On the other hand, when
$L_W\ll v_F/\Omega$, ${\cal G}_{\tilde\Gamma}$ is determined by the
properties of the external leads.  In the RG analysis discussed
earlier, this corresponds to the infrared divergence of $\tilde\Gamma$
being cut off by the energy scale $\hbar v_F/L_W$. Consequently the
charge (spin) $Q^{c (s)}$ is independent of the pumping frequency, and
also of the length $L_W\ll v_F/\Omega$ just as for the non-interacting
case. To see the finite size scaling one can imagine operating the
pump in a closed geometry, for example, in a ring. Then the finite
size effects of $G_{\tilde\Gamma}$ are similar to those for the
ac-conductance in Ref.~\onlinecite{kane92b}.
%

The dependence of $Q$ on the DC conductance also implies that the
finite temperature effects (when $T>\hbar\Omega$) are similar to that
of Ref.~\onlinecite{kane92b}. Consequently the low temperature
($\Omega<T<\bar\omega_\Gamma$) expansion is similar to
(\ref{wire-qsc}) -- when $T$ is substituted for $\Omega$.

In conclusion, let us note here that unlike the case of an applied
voltage \cite{kane92b} the frequency corrections to the pumped charge
are non universal, so that it is not possible to write a one-parameter
(for example, $T_B$) scaling formula for the pumped charge as a
function of frequency. This dependence on the shape of a pumping cycle
also applies for the temperature and finite size effects.


\section{Time-dependent impurity in the Luttinger model at $K=1,\,\frac{1}{2}$}
\label{k-half-sol}

In this section we consider electrons with short ranged
density-density interactions described by the Hamiltonian of a
Luttinger model, written here for the case of spinless electrons:
\begin{equation}
\begin{split}
\label{hll}
H_{LL}&=\int dx\Big\{ \Psi^{\dg}(-i\sigma^3\partial_x)\Psi \\
&+ g_2[\rho_R(x)\rho_R(x)+\rho_L(x)\rho_L(x)]\Big\},
\end{split}
\end{equation}
with the interactions parameterized by $K=\sqrt{(1-g_2)/(1+g_2) }$.
To this we add the backscattering Hamiltonian $H_\Gamma$, introduced earlier.

\subsection{Exact solution for pumped current at $K=1$}
\label{keq-one}
In the absence of interactions, $g_2=0$, the Luttinger parameter
$K=1$. This case can be readily solved for the current. We begin by
writing the Hamiltonian with the time-dependent backscattering
amplitude $\Gamma(t)$:
\begin{eqnarray}
  \label{nonint-h}
  {\cal H}&=&H+H_\Gamma\nonumber\\
&=&\int\! dx\;\Psi^{\dg}(-i\sigma^3\partial_x)\Psi(x,t) \nonumber\\
&+&\delta(x)\left[\Gamma(t)\psi_L^{\dg}\psi_R(x,t) + \Gamma^*(t)\psi_R^{\dg}\psi_L(x,t)\right]
\end{eqnarray}
We remind the reader that we have set $\hbar=1=v_F$. The density of
states per unit length for a single spin species is $\nu_0=1/(\pi\hbar
v_F)\equiv 1/\pi$, and the constraint that the ground state contain a
fixed number of particles relates the upper cut-off of the theory to
the Fermi energy $E_F$. The uniform density of particles in the ground
state of the wire is $k_F/\pi$.  In the absence of any time-dependence
of $\Gamma$ the wire is assumed to be in equilibrium. The current,
which is given by the difference in density of the left and right
movers, is therefore zero. We now ask for the density difference as a
function of time at a point in the wire to the right of the impurity.
For convenience, we choose this point to be in the immediate vicinity
of the impurity. The time evolution of the fields is (obtained by
using the Fermion anticommutation relations):
\begin{subequations}\label{eompsi-rl}
\begin{eqnarray}
-i\partial_t \psi_R(x) &=& [{\cal H}, \psi_R(x)]\nonumber\\
                     &=& i\partial_x \psi_R(x) - \delta(x)\Gamma^*(t)\psi_L(x),    \label{eompsi-r}\\
-i\partial_t \psi_L(x) &=& [{\cal H}, \psi_L(x)]\nonumber\\
    &=& -i\partial_x \psi_L(x) -
\delta(x)\Gamma(t)\psi_R(x) ,    \label{eompsi-l}
\end{eqnarray}
\end{subequations}
Since we are interested in the fields in the immediate vicinity of the
impurity, we integrate these equations to obtain:
\begin{subequations}\label{dpsi-pm}
\begin{eqnarray}
\pmbf{\Delta\psi_R}(t)= -i\Gamma^*(t)\frac{1}{2}\pmbf{\psi_L}(t),
\label{dpsi-r}\\
\pmbf{\Delta\psi_L}(t) = -i\Gamma(t)\frac{1}{2}\pmbf{\psi_R}(t) ,
\label{dpsi-l}
\end{eqnarray}
\end{subequations}
where
\begin{subequations}
\begin{eqnarray}
\pmbf{\Delta\psi_{R}}&=&\psi_{R}(x=0^+)-\psi_{R}(x=0^-),\\
\pmbf{\Delta\psi_{L}}&=&\psi_{L}(x=0^-)-\psi_{L}(x=0^+),\\
\pmbf{\psi_{R,L}}&=&\psi_{R,L}(x=0^+)+\psi_{R,L}(x=0^-).
\end{eqnarray}
\end{subequations}
The field $\psi_R(x,t)$ being a right-mover is a free field to the
left of the impurity, not being influenced by the presence of the
impurity. Likewise for the field $\psi_L(x,t)$ to the right of the
impurity. From the relations (\ref{dpsi-pm}) we can determine the
outgoing fields:
\begin{subequations}\label{psi-out}
\begin{eqnarray}
\psi_R(0^+,t)&=& \frac{-2i\Gamma^*(t)}{1+|\Gamma(t)|^2}\psi_L(0^+,t)\nonumber\\
&+&\frac{1-|\Gamma(t)|^2}{1+|\Gamma(t)|^2}\psi_R(0^-,t) ,    \label{psi-ro}\\
\psi_L(0^-,t)&=& \frac{-2i\Gamma(t)}{1+|\Gamma(t)|^2}\psi_R(0^-,t)
\nonumber\\
&+& \frac{1-|\Gamma(t)|^2}{1+|\Gamma(t)|^2}\psi_L(0^+,t).    \label{psi-lo}
\end{eqnarray}
\end{subequations}
The current for free electrons is given by the
current in the outgoing channel:
\begin{equation}
  \label{ip-ps}
\begin{split}
I(t)&=\frac{1}{k_F}\left\langle\psi_R^{\dg}(0^+,t)
  i\partial_t\psi_R(0^+,t)\right\rangle\\
    & -\frac{1}{k_F}\left\langle\psi_L^{\dg}(0^-,t)
      i\partial_t\psi_L(0^-,t)\right\rangle
    + H.c.,
\end{split}
\end{equation}
as the incoming channel is taken to have zero net current:
$$
\left\langle\psi_L^{\dg}(0^+,t)i\partial_t\psi_L(0^+,t)\right\rangle
-\left\langle\psi_R^{\dg}(0^-,t)i\partial_t\psi_R(0^-,t)\right\rangle=0.
$$
Note that use has been made of Eq.~(\ref{eompsi-rl}) away from
the impurity, to obtain time derivatives in the expression for the
current.

Using Eq.~(\ref{psi-out}), and the incoming state normalization that
\begin{eqnarray}
\left\langle\psi_R^{\dg}(0^-,t)\psi_R(0^-,t)\right\rangle
= \left\langle\psi_L^{\dg}(0^+,t)\psi_L(0^+,t)\right\rangle\nonumber
\equiv \frac{k_F}{2\pi},
\end{eqnarray}
we can write the current (at
zero temperature) as:
\begin{eqnarray}
  \label{ip-1}
  I(t)=\frac{1}{i\pi}\frac{\dot\Gamma(t)\Gamma^*(t)-\dot\Gamma^*(t)\Gamma(t)}
{\left(1+|\Gamma(t)|^2\right)^2}
\end{eqnarray}
The pumped charge (at zero temperature) can now be written as:
\begin{eqnarray}
  \label{qp-k1}
  Q&=&\int dt \frac{1}{i\pi}
  \frac{\dot\Gamma(t)\Gamma^*(t)-\dot\Gamma^*(t)\Gamma(t)}
{\left(1+|\Gamma(t)|^2\right)^2}\\
&=&\frac{1}{2\pi i}\oint \frac{d\Gamma}{\Gamma}
\left[\frac{2|\Gamma|^2}{\left(1+|\Gamma|^2\right)^2}\right] + C.C.
\end{eqnarray}

The reflection probability $|\Gamma|^2$ determines the pumped charge
generated by backscattering of electrons from the barrier. The chief
feature of non-interacting electron gas is the energy independence of
the reflection probability (for low-energies). As a consequence, we
find that the pumped current is determined by the {\it instantaneous}
backscattering amplitude.

\subsubsection{Spin pump }
\label{spinpump-k1}

Including spins in the above analysis, to analyze the spin pump, is
straightforward; different spin species are backscattered by different
amplitudes (see Section \ref{spin-trans}). Therefore, all we have to
do to determine the pumped charge and spin is to put indices of
different spin species on $\Gamma$ in the expression (\ref{ip-1}).
The spin current is then given by:
\begin{eqnarray}
  \label{ispin-k1}
    I_s(t)&=&I_{\up}(t)-I_{\dwn}(t),\\
I_{\up,\dwn}(t)&=&\frac{1}{i\pi}\frac{\dot\Gamma_{\pm}(t)\Gamma_{\pm}^*(t)
-\dot\Gamma_{\pm}^*(t)\Gamma_{\pm}(t)}
{\left(1+|\Gamma_{\pm}(t)|^2\right)^2}
\end{eqnarray}
The condition for generating a pure spin current requires tuning the
amplitudes $\Gamma_\pm$ such that $I_{\up}=-I_{\dwn}$, and is given
by:
$$
\dot\phi_+(t)
\frac{|\Gamma_{+}(t)|^2}{\left(1+|\Gamma_{+}(t)|^2\right)^2}=
-\dot\phi_-(t)
\frac{|\Gamma_{-}(t)|^2}{\left(1+|\Gamma_{-}(t)|^2\right)^2}
$$
where $\phi_\pm(t)$ is the phase of $\Gamma_\pm(t)$. This requires
some fine-tuning of the scattering amplitudes. For example, in the
case of delta function potentials $V_{e,o}(x)$ of Section
\ref{spin-trans}, adjusting the distance ${\ell}$ between the barrier and the
magnetic field to be such that $2k_F\ell=\pi/2$ satisfies this condition.

\subsection{Exact solution for pumped current at $K=1/2$}
\label{sec-khalf}

We now consider the case with non-zero interactions ($g_2\neq 0$) in
the Luttinger model for spinless electrons.  With the bulk Hamiltonian
$H=H_{LL}$ at $K=1/2$, the backscattering Hamiltonian $H_{\Gamma}$ of
Eq.~(\ref{himp}) has a scaling dimension $\Delta=K=1/2$ -- same as
that of a Fermi field operator. The low energy behavior of the theory
with the impurity is therefore identical to that of a system of chiral
Fermions (different from the original interacting fermions) with a
Hamiltonian \cite{guinea85,matveev95,chamon95}:
\begin{equation}\label{Hpsi}
\begin{split}
{\cal H} = H+H_{\Gamma}=&\int dx\Big \{\psi^\dagger(x)\left[-i\partial_x \right]\psi(x)\Big \}\\
+& \frac{i}{\sqrt{2}}\left[\lambda \;\hat a\; \psi(0) + \lambda^*\;\hat a \;\psi^\dagger(0)\right],
\end{split}
\end{equation}
where $\psi$ is a chiral Dirac Fermion and $\hat a$ is a Majorana
Fermion representing the impurity. The impurity potential
$\lambda=(1/\sqrt{\pi\alpha})\Gamma$, where $\Lambda_0\equiv
1/\pi\alpha\lesssim E_F$ is the high energy cut-off of the bulk
Hamiltonian $H$, and we recognize the dimensionless coupling $\Gamma$
from the notations used earlier.  In this section we use the mapping
to free fermions to find the exact expression for the non-equilibrium
current arising due to the pumping parameter $\Gamma\to\Gamma(t)$
acquiring a time dependence. This expression was first used in our
earlier publication, Ref.~\onlinecite{sharma01}. Here we show the
details of our calculations for the charge transported across the wire
in a cycle.  In the asymptotically slow limit of pumping (frequency
$\Omega/E_F\to 0$) the pumped charge is shown to be determined by the
dc conductance of the system. Furthermore, a general scaling formula
for the pumped charge is also conjectured.

We begin by defining chiral Majorana fermions
\begin{equation}\label{psicom}
\begin{split}
\eta_1(x)=&\frac{\psi(x)+\psi^{\dg}(x)}{\sqrt{2}};\,\,\eta_2(x)=\frac{\psi(x)-\psi^{\dg}(x)}{i\sqrt{2}}\\
\{ \eta_j(x),&\, \eta_j(x')\}=\{ \psi (x), \psi^\dg (x') \}= \delta (x - x') ;\\
\{ \eta_1(x),&\, \eta_2(x')\}=0\;;\{ \eta_j(x), \hat a \}=0\; ; \{\hat a,\hat a \}=1,
\end{split}
\end{equation}
and denote the real and imaginary parts of the complex scattering matrix $\lambda(t)$ as:
\begin{equation}\label{lambdas}
{\rm Re}\lambda=\lambda_1;\,{\rm Im}\lambda=-\lambda_2;\,\,|\lambda|=\lambda_m.
\end{equation}
Next, we write the  equations of motion for these fields:
\begin{subequations}
\begin{eqnarray}
-i\partial_t \eta_1(x) = [{\cal H}, \eta_1(x)]
                     = i\partial_x \eta_1(x) +
i\lambda_1\delta(x) \hat a ,    \label{eompsi}\\
-i\partial_t \eta_2(x) = [{\cal H}, \eta_2(x)]
    = i\partial_x \eta_2(x) +
i\lambda_2 \delta(x)\hat a ,    \label{eompsid}\\
i\partial_t \hat a =- [{\cal H}, \hat a]
              = i\lambda_1 \eta_1(0) +
i\lambda_2\eta_2(0).    \label{eomf}
\end{eqnarray}
\end{subequations}
Now we note that the Majorana fields are right-movers and
therefore, the fields to the right of the impurity are dependent on the
fields to its left. The latter are, however, independent of the impurity
potential.

To find how the chiral Majorana fermion fields at $x=0^+$ are related
to those at $x=0^-$, we integrate across the impurity, and using the
notation
$$ \Delta\eta_j =\frac{1}{2}[\eta_j(0^+)-\eta_j(0^-)],$$
obtain:
\begin{equation}
  \label{eom-a1}
  \begin{split}
-\Delta\eta_{1}&= \frac{1}{2}\lambda_1\hat a;\,\,\,
-\Delta\eta_{2} = \frac{1}{2}\lambda_2\hat a;\\
\partial_t\hat a&=\lambda_1[\Delta\eta_{1} +{\eta_{1}}(0^-)]
+\lambda_2[\Delta\eta_{2} +{\eta_{2}}(0^-)].
\end{split}
\end{equation}
We will now define {\it{scaled}} fields:
$$\eta_1\to \lambda_1^{-1}\eta_1,\,\eta_2\to\lambda_2^{-1}\eta_2 ,$$
and eliminate $\hat a$ between the equations in (\ref{eom-a1}) to obtain:
\begin{equation}
  \label{eom-a2}
\begin{split}
-\partial_t\Delta\eta_{1}&= \frac{1}{2}\Big[\lambda_1^2 \left(\Delta\eta_{1}
+\eta_{1}(0^-) \right) + (1\to 2) \Big]\\
&=-\partial_t\Delta\eta_{2}.
\end{split}
\end{equation}
Since the scaled fields satisfy $\Delta\eta_1=\Delta\eta_2$,
we choose an {\it ansatz} for relating the fields across the impurity:
\begin{equation}
  \label{ansatz}
\begin{split}
\Delta\eta_{1}&=\bar M_{1}* {\eta_{1}}
+ \bar M_{2}* {\eta_{2}}
=\Delta\eta_{2}
\end{split}
\end{equation}
Here $*$ denotes the convolution:
$$\bar M_{1}* {\eta_{1}}=\int_{-\infty}^t dt_1\bar M_{1}(t;t_1){\eta_{1}}(0^-, t_1).$$
From Eq.~(\ref{eom-a2}) we get the following differential equations for the $M's$:
\begin{equation}
  \label{deq-a}
  \begin{split}
\partial_t\bar M_{j}(t;t_1) &+
\frac{1}{2}\left[\lambda_1^2+\lambda_2^2\right]\bar M_{j}(t;t_1)\\
&+\frac{1}{2}\lambda_{j}^2\delta(t-t_1)=0;\,\,{\rm for}\,j=1,2
\end{split}
\end{equation}
These can be solved straightforwardly to reveal:
\begin{subequations}
\begin{equation}
  \label{m-tt1}
\begin{split}
\bar M_{j}(t;t_1)&=-\frac{\lambda^2_{j}(t_1)}{2}\theta(t-t_1)\\
&\times
{\rm exp}\left\{-\int_{t_1}^{t}dt'\frac{1}{2}\left[\lambda_1^2(t')+\lambda_2^2(t')\right]\right\}.
\end{split}
\end{equation}
We thus find the relation between the {\it scaled} fields on either
sides of the impurity:
\begin{equation}
  \label{etapm-a}
  \begin{split}
{\eta_{j}(0^+, t)}&= {\eta_{j}}(0^-, t) + \sum_{k=1}^{2}2\bar M_{k}*{\eta_{k}}
\end{split}
\end{equation}
\end{subequations}

Reverting to original {\it unscaled} fields we define the kernels:
\begin{eqnarray}
  \label{mn}
M_j(t;t_1)&=& 2\bar M_{j}(t;t_1)\lambda^{-1}_{j}(t_1),
\end{eqnarray}
and write the relations between Majorana fermions across $x=0$:
\begin{subequations}
 \label{mf-pm}
\begin{eqnarray}
{\eta_{1}}(0^+,t)= {\eta_{1}}(0^-,t)
+ \lambda_1(t)\Big[M_{1}*{\eta_{1}}
+ M_{2}*{\eta_{2}}\Big]\label{mf-pm1}\\
{\eta_{2}}(0^+,t)= {\eta_{2}}(0^-,t)
+ \lambda_2(t)\Big[M_{1}*{\eta^{1}_{\rm f}}
+ M_{2}*{\eta^{2}_{\rm f}}\Big]\label{mf-pm2}
\end{eqnarray}
\end{subequations}
From the equations (\ref{mf-pm}) above we get the relations in terms of the Dirac fermions:
\begin{equation}
  \label{psi-pm}
  \begin{split}
\psi(0^+,t) &=\psi(0^-,t)
+ \lambda^*(t)\Big\{ M_{1}*\left[\psi +\psi^{\dg}\right]\\
&+ i M_{2}*\left[\psi^{\dg} - \psi\right]\Big\}.
\end{split}
\end{equation}
The above equation allows us to calculate the current at time $t$. The
current is given by the difference in the electron density across the
impurity
\begin{equation}
  \label{psi-pm-i}
  \begin{split}
I(t)=\left\langle\psi^{\dg}(0^+,t)\psi(0^+,t)  - \psi^{\dg}(0^-,t) \psi(0^-,t)\right\rangle,
\end{split}
\end{equation}
where the expectation value is taken in the Heisenberg state at a time
in the remote past when the system is in equilibrium. Using:
\begin{equation}
\begin{split}
\left\langle\psi^{\dg}(0^-,t)\psi(0^-,t')\right\rangle &=
\left\langle\psi(0^-,t)\psi^{\dg}(0^-,t')\right\rangle\\
=D(t-t')&=\sum_E n(E)e^{iE(t - t')},
\end{split}
\end{equation}
 we find:
\begin{equation}
  \label{i-phidot}
 \begin{split}
I(t)=&\int_{-\infty}^{t}\! dt'\;\dot\phi(t')\cos[\phi(t)-\phi(t')]\\
&\times\int_{-\infty}^{t'}dt'' K^{(R)}(t,t'').
\end{split}
\end{equation}
where the retarded current correlator
\begin{equation}
  \label{kr}
 \begin{split}
K^{(R)}(t,t'')=\lambda_m(t)&\lambda_m(t'')
{\rm exp}\Big[\frac{1}{2}\int_{t}^{t''}dt'''\lambda_m^2(t''')\Big]\\
&\times\int_{-\Lambda_0}^{\Lambda_0}\frac{dE}{2\pi} n(E) \sin E(t -t'')
\end{split}
\end{equation}

\subsection{Pumped charge: low frequency asymptotics}
\label{khalf-asymptotics}

Consider the pumping cycle in a time interval $[t_i,t_f]$, so that
$\dot\phi(t)=0, \,\,t<t_i$.  The expression (\ref{i-phidot}) above can
be integrated over time to find the charge pumped in the cycle:
\begin{equation}\label{qp-half}
\begin{split}
Q=\int_{t_i}^{t_f}dt'\dot\phi(t')\int_{t'}^{t_f}&dt\cos\left[\phi(t)-\phi(t')\right]\\
\times&\int^{t'}_{-\infty}dt''K^{(R)}(t;t'')\\
\equiv &\int_{t_i}^{t_f}dt'\dot\phi(t'){\cal G}(t'),
\end{split}
\end{equation}

We now seek the asymptotic behavior of ${\cal G}$ as the ratio
$\Omega/\Lambda_0\to 0$, where $\Omega$ is the pumping frequency and
$\Lambda_0\lesssim E_F$ is the upper cut-off of the field theory
(\ref{Hpsi}).  Consider first the case of zero temperature. Then the energy integral can
be performed easily and we obtain:
\begin{eqnarray}
  \label{kr-sing}
  K^{(R)}(t,t'')&=&-\lambda_m(t)\lambda_m(t'')e^{\frac{1}{2}\int_{t}^{t''}d\tilde t\lambda_m^2(\tilde t)}
\nonumber\\
&\times&
\left[\frac{1-\cos\Lambda_0(t-t'')}{2\pi(t-t'')}\right]
\end{eqnarray}
This function has a maximum at $t-t''\approx 1/\Lambda_0$, with a
width $\delta t^*\approx 1/\omega_\Gamma + 1/\Lambda_0\sim
1/\omega_\Gamma$, where $\omega_\Gamma=\lambda_m^2(t')\ll\Lambda_0$.
As a result, the time $t'$ has to be within $1/\Lambda_0$ of this
maximum, and we can write: $\lambda_m(\Omega t)\approx\lambda_m(\Omega
t')$.  We note that the neglected term, $\delta\lambda_m/\lambda_m$,
contributes only to even orders to the current, {\it i.e.},
$(\delta\lambda_m/\lambda_m)^{2n}\sim(\Omega/\omega_\Gamma)^{2n}$, for
$n=1,2\dots$. It can therefore be accounted for by a suitable renormalization
of the energy scale $\omega_\Gamma$. We therefore focus on the analytics of the lowest
order corrections, and write:
\begin{equation}\label{ical-1}
\begin{split}
{\cal G}(t')&=\int_{0}^{\infty}d t \int_{-\infty}^{0}dt''
\lambda_m^2(t') e^{-\frac{1}{2}\lambda_m^2(t')(t-t'')}\\
&\times\int_{-\Lambda_0}^{\Lambda_0}\frac{dE}{2\pi}
n(E)\sin[E(t-t'')].
\end{split}
\end{equation}
Using the time-translation invariance of the resulting integrand, we have:
\begin{equation}\label{ical-2}
\begin{split}
  {\cal G}(t')& =\int_{-\Lambda_0}^{\Lambda_0}\frac{dE}{2\pi} n(E)
  \int_{-\infty}^{0}d\tau\;\tau\,\omega_\Gamma
  e^{\frac{1}{2}\omega_\Gamma\,\tau}
  \sin[E \tau]\\
&=\frac{1}{\pi}-
\Big[\frac{1}{\pi}-2\int_{-\Lambda_0}^{\Lambda_0}
\frac{dE}{2\pi}\frac{dn(E)}{dE}
\frac{\omega_\Gamma^2(t')}{\omega_\Gamma^2(t')+4E^2}\Big]\\
&=\frac{1}{\pi}-2 G(\Omega\to 0),
\end{split}
\end{equation}
where $G(\Omega\to 0)$ is the dc conductance of the $K=1/2$ Luttinger
liquid \cite{matveev95,fendley95} with an impurity of strength
$\omega_\Gamma^2$. It has a low temperature ($T<\omega_\Gamma$)
behavior $G\sim (T/\frac{1}{2}\lambda_m^2(t'))^2$.  This implies that
the pumped charge:
\begin{equation}
  \label{q-slowpumping}
 \begin{split}
Q=\frac{1}{2\pi}\int dt'\dot\phi(t')\left[1-2\pi G(\Omega\to 0)\right],
\end{split}
\end{equation}
at zero temperature, is $Q=e$.


\subsection{Crossover temperature and scaling behavior}
\label{tb-mspt}


In order to look at the high frequency pumping limit, we consider the
time dependence of $\phi$ and $\lambda_m$ with a principle frequency
$\Omega$, so that we can write:
\begin{equation}\label{recale}
\lambda_m(t)\equiv\lambda_m(\Omega t),\,\,\,\phi(t)\equiv\phi(\Omega t).
\end{equation}
Then it is straightforward to see that the expression for pumped
charge (\ref{qp-half}) can be written so as to make all the $\Omega$
dependence explicit (upon rescaling all the times by a factor
$\Omega$):
\begin{equation}\label{qp-scaled}
\begin{split}
Q=\int_{0}^{2\pi}dt'\dot\phi(t')\int_{t'}^{2\pi}&dt\cos\left[\phi(t)-\phi(t')\right]\\
\times&\int^{t'}_{-\infty}dt''\frac{\lambda_m(t)\lambda_m(t'')}{\Omega}
e^{\frac{1}{2}\int_{t}^{t''}dt_1\lambda_m^2(t_1)/\Omega}\\
&\times\int \frac{dE}{2\pi} \tilde n(E) \sin E(t -t'')
\end{split}
\end{equation}
From this expression, a large-$\Omega/\omega_\Gamma$ expansion follows
when the argument of the exponential is a small quantity, {\it i.e.}, when
$$\frac{\int_{0}^{2\pi}dt_1{\lambda_m}^2(t_1)}{\Omega}\ll 1,$$
where $\lambda_m(t)$ has its times rescaled according to (\ref{recale}).
Expanding the exponential we can compare the resulting expansion with
the low-frequency expansion of (\ref{qpump-lf}), by substituting for
$K\to 1/K$.  We thus find the coefficients $\{q_n\}$ for all integer
powers $n$ of $1/\Omega$ in this {\it dual} expansion. It is clear that
unless $\lambda_m$ is time-independent, the ratios of these coefficients
are not universal numbers.

\begin{figure}
\includegraphics[scale=1.1]{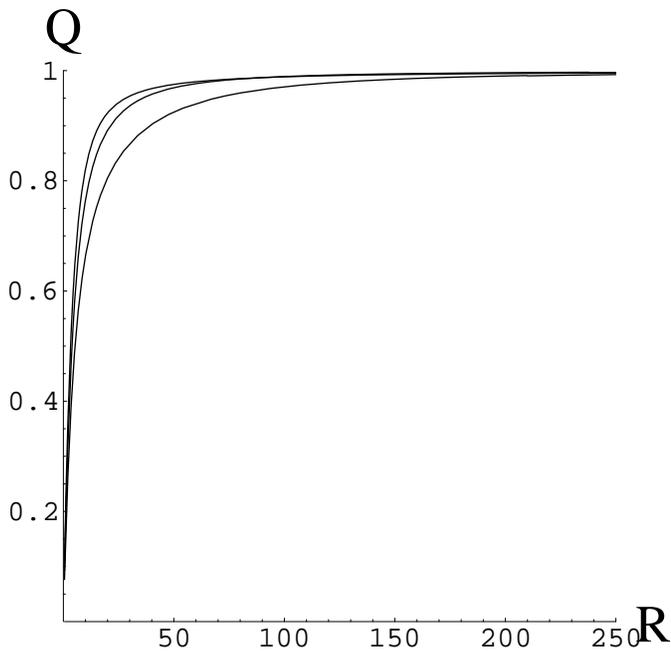}
\caption{Variation of pumped charge with frequency for three different
  shapes of the pumping cycle}
\label{khalf-plot}
\end{figure}

A particular form of time-dependence for the parameters $\phi(t)$ and $\lambda_m(t)$
based upon the simple model of Section \ref{special-form} is:
\begin{eqnarray}
  \label{func-form}
  \lambda_m^2(t)&=& \lambda_0^2\left[1 + \dl\cos2\Omega t\right],\,\,\,0\leq \dl<1\\
 \phi(t)&=&
\tan^{-1}\left[\frac{\sqrt{1+\dl}\cos\Omega t  - \sqrt{1-\dl}\sin\Omega t }
{ \sqrt{1+\dl}\cos\Omega t  +\sqrt{1-\dl}\sin\Omega t }\right]\\
\dot\phi(t)&=&\frac{\sqrt{1-\dl^2}}{1+\dl\cos2\Omega t}
\end{eqnarray}
For $\dl=0$ this gives a constant $\dot\phi$ and so we can compute the
charge pumped for any frequency $\Omega$ in terms of known functions.
We find, after straightforward but tedious algebra:
\begin{equation}
  \label{q-dc}
 \begin{split}
Q=\frac{1}{2\pi}\int dt'\dot\phi(t')\left[1-2\pi G(\Omega)\right],
\end{split}
\end{equation}
where $G(\Omega)$ is the conductance at a dc voltage $\Omega$ for the $K=1/2$ Luttinger model
\cite{matveev95,fendley95}
and at zero temperature is given by:
$$
G(\Omega)=\frac{1}{2\pi}-\frac{\lambda_0^2}{2\pi\Omega}\tan^{-1}
\left[\frac{\Omega}{\lambda_0^2}\right].
$$
In Fig.~\ref{khalf-plot} the pumped charge in a cycle $Q$ is plotted
versus $R\equiv\frac{\lambda_0^2}{2\Omega}\equiv
\frac{\omega_\Gamma}{\Omega}$ for three values of the parameter
$\dl=0,0.5,0.7$. It is clear from the plots that the low and high
frequency asymptotic behavior is the same in all three cases. The
uppermost curve is for $\dl=0$ -- a constant $\lambda_m$ and therefore
a circular pumping cycle -- and has the largest pumped charge $Q$ per
cycle for any particular value of $R$. This decrease in pumped charge
for a given frequency is to be expected, as an oscillatory amplitude
leads the response function, ${\cal G}$, out of phase with the driving
force $\dot\phi$, thereby decreasing the current.  As $R$ increases
from $0$, deviations from $\dl=0$ curve are apparent near $R\sim 10$;
at low frequency (large $R$) the curves for increasing $\dl$
($\dl=0.5$ and $\dl=0.7$ in the figure) converge with the curve for
$\dl=0$ at progressively greater values of $R$, indicating the
decoherent (non-adiabatic) nature of deviations from the circular
shape of the pumping cycle.

Thus, although there is a range of frequency where pumped charge is
dependent on all the details of the time-dependent parameters in the
problem, the asymptotic behavior is universal. A scaling formula for
the pumped charge can therefore be written when $\Gamma(t)\equiv
\Gamma(\Omega t)$:
\begin{equation}
  \label{scaling-form}
  Q=\tilde Q(\Omega/\omega_\Gamma,\delta\omega_\Gamma/\omega_\Gamma)
\end{equation}
where $\omega_\Gamma=\frac{1}{2}\int_{0}^{2\pi/\Omega}dt\lambda_m^2(t)$
is a cross-over energy scale in the time-dependent problem, and
$\delta\omega_\Gamma$ represents a collection of other
(time-independent) parameters contained in $\Gamma$ similar to the
quantity $\dl$ in Eq.~(\ref{func-form}).  The behavior of the scaling
function is described by:
$$
\tilde Q(0,x)=1,\,\,\tilde Q(x,y)\leq \tilde Q(x,0).
$$
Let us note that it has not been proven that the inequality will
hold for a more general form for $\lambda(t)$.

\section{Conclusion}

In conclusion, we have studied the adiabatic limit of a quantum pump
in a quantum wire, by mapping it to a problem of a time-dependent
backscatterer in a Luttinger liquid. We have shown that the properties
of scale invariance and chiral symmetry of the Luttinger liquid imply
a relationship between the pumped quantities and the dc conductance.
The difference between a Fermi liquid and a Luttinger liquid behavior
in a one-dimensional system is clearly brought out by this mechanism
of charge and spin transport. It is found that for a Fermi liquid the
adiabatic regime -- for which the pumped quantities per cycle are
independent of pumping frequency -- extends up to pumping frequency
$\Omega<E_F$, whereas for the Luttinger liquid this regime of pumping
is limited to $\Omega\ll\omega_\Gamma\ll E_F$. The barrier energy
scale $\omega_\Gamma$ is found to be a crossover energy scale between
adiabatic and impulsive (sudden) response to the time-dependence. This
distinction is absent in the Fermi liquid picture. Another way of
interpreting these results is to associate adiabaticity, for this
mechanism of pumping, with the scale invariant fixed points of a wire
with an impurity.  For the non-interacting case the {\it beta}
function for the impurity coupling is zero, consequently any pumping
frequency implies an adiabatic response. For the interacting Luttinger
liquid the low-energy fixed point corresponds to an infinite barrier.
Therefore, the adiabaticity criterion necessarily depends on the ratio
$\Omega/\omega_\Gamma$.

\begin{acknowledgements}
We thank Ian Affleck and Paul Krapivsky
for several helpful suggestions. This work was supported by the NSF
under grant No.~DMR-98-76208 (P.~S. and C.~C.)
and the Alfred P.~Sloan Foundation (C.~C.).
\end{acknowledgements}

\begin{thebibliography}{23}
\expandafter\ifx\csname natexlab\endcsname\relax\def\natexlab#1{#1}\fi
\expandafter\ifx\csname bibnamefont\endcsname\relax
  \def\bibnamefont#1{#1}\fi
\expandafter\ifx\csname bibfnamefont\endcsname\relax
  \def\bibfnamefont#1{#1}\fi
\expandafter\ifx\csname citenamefont\endcsname\relax
  \def\citenamefont#1{#1}\fi
\expandafter\ifx\csname url\endcsname\relax
  \def\url#1{\texttt{#1}}\fi
\expandafter\ifx\csname urlprefix\endcsname\relax\def\urlprefix{URL }\fi
\providecommand{\bibinfo}[2]{#2}
\providecommand{\eprint}[2][]{\url{#2}}

\bibitem[{\citenamefont{Thouless}(1983)}]{thouless83}
\bibinfo{author}{\bibfnamefont{D.~J.} \bibnamefont{Thouless}},
  \bibinfo{journal}{Phys. Rev. B} \textbf{\bibinfo{volume}{27}},
  \bibinfo{pages}{6083} (\bibinfo{year}{1983}).

\bibitem[{\citenamefont{Niu}(1986)}]{niu86}
\bibinfo{author}{\bibfnamefont{Q.}~\bibnamefont{Niu}}, \bibinfo{journal}{Phys.
  Rev. B} \textbf{\bibinfo{volume}{34}}, \bibinfo{pages}{5093}
  (\bibinfo{year}{1986}).

\bibitem[{\citenamefont{Niu}(1990)}]{niu90}
\bibinfo{author}{\bibfnamefont{Q.}~\bibnamefont{Niu}}, \bibinfo{journal}{Phys.
  Rev. Lett.} \textbf{\bibinfo{volume}{64}}, \bibinfo{pages}{1812}
  (\bibinfo{year}{1990}).

\bibitem[{\citenamefont{Brouwer}(1998)}]{brouwer98}
\bibinfo{author}{\bibfnamefont{P.~W.} \bibnamefont{Brouwer}},
  \bibinfo{journal}{Phys. Rev. B} \textbf{\bibinfo{volume}{58}},
  \bibinfo{pages}{10135} (\bibinfo{year}{1998}).

\bibitem[{\citenamefont{Aleiner and Andreev}(1998)}]{aleiner98}
\bibinfo{author}{\bibfnamefont{I.~L.} \bibnamefont{Aleiner}} \bibnamefont{and}
  \bibinfo{author}{\bibfnamefont{A.~V.} \bibnamefont{Andreev}},
  \bibinfo{journal}{Phys. Rev. Lett.} \textbf{\bibinfo{volume}{81}},
  \bibinfo{pages}{1286} (\bibinfo{year}{1998}).

\bibitem[{\citenamefont{Zhou et~al.}(1999)\citenamefont{Zhou, Spivak, and
  Altshuler}}]{zhou99}
\bibinfo{author}{\bibfnamefont{F.}~\bibnamefont{Zhou}},
  \bibinfo{author}{\bibfnamefont{B.}~\bibnamefont{Spivak}}, \bibnamefont{and}
  \bibinfo{author}{\bibfnamefont{B.}~\bibnamefont{Altshuler}},
  \bibinfo{journal}{Phys. Rev. Lett.} \textbf{\bibinfo{volume}{82}},
  \bibinfo{pages}{608} (\bibinfo{year}{1999}).

\bibitem[{\citenamefont{Sharma and Chamon}(2001)}]{sharma01}
\bibinfo{author}{\bibfnamefont{P.}~\bibnamefont{Sharma}} \bibnamefont{and}
  \bibinfo{author}{\bibfnamefont{C.}~\bibnamefont{Chamon}},
  \bibinfo{journal}{Phys. Rev. Lett.} \textbf{\bibinfo{volume}{87}},
  \bibinfo{pages}{096401} (\bibinfo{year}{2001}).

\bibitem[{\citenamefont{Feldman and Gefen}(2001)}]{feldman-comment}
\bibinfo{author}{\bibfnamefont{D.~E.} \bibnamefont{Feldman}} \bibnamefont{and}
  \bibinfo{author}{\bibfnamefont{Y.}~\bibnamefont{Gefen}},
  \bibinfo{journal}{cond-mat/0111411}  (\bibinfo{year}{2001}).

\bibitem[{\citenamefont{Stone}(1994)}]{stone}
\bibinfo{author}{\bibfnamefont{M.}~\bibnamefont{Stone}},
  \emph{\bibinfo{title}{Bosonization}} (\bibinfo{publisher}{World Scientific},
  \bibinfo{year}{1994}).

\bibitem[{\citenamefont{Chamon et~al.}(1995)\citenamefont{Chamon, Freed, and
  Wen}}]{chamon95}
\bibinfo{author}{\bibfnamefont{C.}~\bibnamefont{Chamon}},
  \bibinfo{author}{\bibfnamefont{D.~E.} \bibnamefont{Freed}}, \bibnamefont{and}
  \bibinfo{author}{\bibfnamefont{X.~G.} \bibnamefont{Wen}},
  \bibinfo{journal}{Phys. Rev. B} \textbf{\bibinfo{volume}{51}},
  \bibinfo{pages}{2363} (\bibinfo{year}{1995}).

\bibitem[{\citenamefont{Schwinger}(1961)}]{schwinger61}
\bibinfo{author}{\bibfnamefont{J.}~\bibnamefont{Schwinger}},
  \bibinfo{journal}{J. Math. Phys. (N.Y.)} \textbf{\bibinfo{volume}{2}},
  \bibinfo{pages}{407} (\bibinfo{year}{1961}).

\bibitem[{\citenamefont{Rammer and Smith}(1986)}]{rammer86}
\bibinfo{author}{\bibfnamefont{J.}~\bibnamefont{Rammer}} \bibnamefont{and}
  \bibinfo{author}{\bibfnamefont{H.}~\bibnamefont{Smith}},
  \bibinfo{journal}{Rev. Mod. Phys.} \textbf{\bibinfo{volume}{58}},
  \bibinfo{pages}{323} (\bibinfo{year}{1986}).

\bibitem[{\citenamefont{Zinn-Justin}(1996)}]{zinnjustin}
\bibinfo{author}{\bibfnamefont{J.}~\bibnamefont{Zinn-Justin}},
  \emph{\bibinfo{title}{Quantum field theory and critical phenomena}}
  (\bibinfo{publisher}{Oxford University Press}, \bibinfo{year}{1996}).

\bibitem[{\citenamefont{Kane and Fisher}(1992)}]{kane92b}
\bibinfo{author}{\bibfnamefont{C.~L.} \bibnamefont{Kane}} \bibnamefont{and}
  \bibinfo{author}{\bibfnamefont{M.~P.~A.} \bibnamefont{Fisher}},
  \bibinfo{journal}{Phys. Rev. B} \textbf{\bibinfo{volume}{46}},
  \bibinfo{pages}{15233} (\bibinfo{year}{1992}).

\bibitem[{\citenamefont{Furusaki and Nagaosa}(1996)}]{furusaki96}
\bibinfo{author}{\bibfnamefont{A.}~\bibnamefont{Furusaki}} \bibnamefont{and}
  \bibinfo{author}{\bibfnamefont{N.}~\bibnamefont{Nagaosa}},
  \bibinfo{journal}{Phys. Rev. B} \textbf{\bibinfo{volume}{54}},
  \bibinfo{pages}{5239} (\bibinfo{year}{1996}), \bibinfo{note}{and references
  therein}.

\bibitem[{\citenamefont{Fendley et~al.}(1995)\citenamefont{Fendley, Ludwig, and
  Saleur}}]{fendley95}
\bibinfo{author}{\bibfnamefont{P.}~\bibnamefont{Fendley}},
  \bibinfo{author}{\bibfnamefont{A.~W.} \bibnamefont{Ludwig}},
  \bibnamefont{and} \bibinfo{author}{\bibfnamefont{H.}~\bibnamefont{Saleur}},
  \bibinfo{journal}{Phys. Rev. B} \textbf{\bibinfo{volume}{52}},
  \bibinfo{pages}{8934} (\bibinfo{year}{1995}).

\bibitem[{\citenamefont{Wen}(1990)}]{wen90}
\bibinfo{author}{\bibfnamefont{X.~G.} \bibnamefont{Wen}},
  \bibinfo{journal}{Phys. Rev. B} \textbf{\bibinfo{volume}{41}},
  \bibinfo{pages}{12838} (\bibinfo{year}{1990}).

\bibitem[{\citenamefont{Wen}(1992)}]{wen92}
\bibinfo{author}{\bibfnamefont{X.~G.} \bibnamefont{Wen}},
  \bibinfo{journal}{Int. J. Mod. Phys. B} \textbf{\bibinfo{volume}{6}},
  \bibinfo{pages}{1711} (\bibinfo{year}{1992}).

\bibitem[{\citenamefont{Laughlin}(1987)}]{Laughlin-in-Prange&Girvin}
\bibinfo{author}{\bibfnamefont{R.~B.} \bibnamefont{Laughlin}}, in
  \emph{\bibinfo{booktitle}{The Quantum Hall Effect}}, edited by
  \bibinfo{editor}{\bibfnamefont{R.~E.} \bibnamefont{Prange}} \bibnamefont{and}
  \bibinfo{editor}{\bibfnamefont{S.~M.} \bibnamefont{Girvin}}
  (\bibinfo{publisher}{New York: Springer-Verlag}, \bibinfo{year}{1987}).

\bibitem[{\citenamefont{Lesage et~al.}(1997)\citenamefont{Lesage, Saleur, and
  Simonetti}}]{lesage97}
\bibinfo{author}{\bibfnamefont{F.}~\bibnamefont{Lesage}},
  \bibinfo{author}{\bibfnamefont{H.}~\bibnamefont{Saleur}}, \bibnamefont{and}
  \bibinfo{author}{\bibfnamefont{P.}~\bibnamefont{Simonetti}},
  \bibinfo{journal}{Phys. Rev. B} \textbf{\bibinfo{volume}{56}},
  \bibinfo{pages}{7598} (\bibinfo{year}{1997}).

\bibitem[{\citenamefont{Guinea et~al.}(1985)\citenamefont{Guinea, Hakim, and
  Muramatsu}}]{guinea85}
\bibinfo{author}{\bibfnamefont{F.}~\bibnamefont{Guinea}},
  \bibinfo{author}{\bibfnamefont{V.}~\bibnamefont{Hakim}}, \bibnamefont{and}
  \bibinfo{author}{\bibfnamefont{A.}~\bibnamefont{Muramatsu}},
  \bibinfo{journal}{Phys. Rev. Lett.} \textbf{\bibinfo{volume}{54}},
  \bibinfo{pages}{263} (\bibinfo{year}{1985}).

\bibitem[{\citenamefont{Matveev}(1995)}]{matveev95}
\bibinfo{author}{\bibfnamefont{K.~A.} \bibnamefont{Matveev}},
  \bibinfo{journal}{Phys. Rev. B} \textbf{\bibinfo{volume}{51}},
  \bibinfo{pages}{1743} (\bibinfo{year}{1995}).

\bibitem[{\citenamefont{Haldane}(1981)}]{haldane81}
\bibinfo{author}{\bibfnamefont{F.~D.~M.} \bibnamefont{Haldane}},
  \bibinfo{journal}{J. Phys. C} \textbf{\bibinfo{volume}{14}},
  \bibinfo{pages}{2585} (\bibinfo{year}{1981}).

\end{thebibliography}

\end{document}